\documentclass[aps,prb,eqsecnum,amsmath,floatfix,showpacs,amssymb,superscriptaddress,twocolumn,10pt]{revtex4-1}
\usepackage[dvips]{color}
\usepackage{graphicx}
\usepackage{float,rotating}
\usepackage[caption=false]{subfig}
\begin{document}
\title{Effects of colored noise on Landau-Zener Transitions: Two and Three-Level Systems}
\author{M. B. Kenmoe}
\affiliation{Mesoscopic and Multilayer Structures Laboratory, Faculty of Science, Department of Physics, University of Dschang, Cameroon}
\affiliation{The Abdus Salam International Centre for Theoretical Physics, 34151 Trieste, Italy}
\author{H. N. Phien}
\affiliation{Center for Engineered Quantum Systems, School of Mathematics and Physics, University of Queensland, Brisbane 4072, Australia}
\author{M. N. Kiselev}
\affiliation{The Abdus Salam International Centre for Theoretical Physics, 34151 Trieste, Italy}
\author{L. C. Fai}
\affiliation{Mesoscopic and Multilayer Structures Laboratory, Faculty of Science, Department of Physics, University of Dschang, Cameroon}
\affiliation{The Abdus Salam International Centre for Theoretical Physics, 34151 Trieste, Italy}
\affiliation{Higher Teachers' Training College, Department of Physics, The University of Bamenda, Cameroon}
\date{\today}
\begin{abstract}
We investigate the Landau-Zener transition in two- and three- level systems subject to a classical Gaussian noise.
Two complementary limits of the noise being fast and slow compared to characteristic Landau-Zener tunnel times are
discussed. The analytical solution of a density matrix (Bloch) equation is given for a long-time asymptotic of
transition probability. It is demonstrated that the transition probability induced or assisted by the fast noise can be
obtained through a procedure of {\it Bloch's equation averaging} with further reducing it to a master equation.
In contrast to the case of fast noise, the transition probability for LZ transition induced or assisted by the slow
classical noise can be obtained by averaging the  {\it solution} of Bloch's equation over the noise realization.
As a result, the transition probability is described by the activation Arrhenius law.  The approximate solution of
the Bloch's equation  at finite times is written in terms of Fresnel's integrals and interpreted in terms of interference
pattern. We discuss consequences of a local isomorphism between $SU(2)$ and $SO(3)$ groups and
connections between Schr\"odinger and Bloch descriptions of spin dynamics. Based on this isomorphism  we establish the relations
between $S=1/2$ and $S=1$ transition probabilities influenced by the noise. A possibility to use the slow noise as a probe for 
tunnel time is discussed.
\end{abstract}
\maketitle
\section{Introduction}\label{Sec1}

The interest to Landau-Zener (LZ) model\cite{lan, zen, stu, Majorana} is constantly increasing
over the last decades as it establishes one of the most important fundamental notions in non-stationary quantum
mechanics. The original LZ model describes the probability of transition between two quantum isolated states coupled by
a linearly sweeping external transverse field of a constant amplitude and a time-dependent longitudinal field  that
passes through
resonance with the transition frequency. Even though the resulting LZ formula seems to be quite simple, it has proved to
be applicable in analyzing the experimental data on
charge transfer particle collisions\cite{prigogine}. The model has been employed in various studies related to charge
transport in nanostructures
\cite{sessoli1999.1, Averin1995, Averin1999, Gefen}, Bose-Einstein condensates\cite{fishman, Renzoni, Armour}, spin tunneling
of nanomagnets\cite{Sessoli2000, Sessoli2002} and quantum quenches\cite{Silva2011, Atland, Altman}. Especially,
LZ becomes a corner stone of theories discussing adiabatic quantum computing\cite{Farhi, Aharonov, Nazarov}
due to a possibility to  enhance a read out of qubits via the Zener flip tunneling\cite{Gradbert}.
Such a mechanism has previously been implemented for flux qubits\cite{collin} and may serve also for inverting spin
population by sweeping the system through the resonance (rapid passage) in ultracold molecules\cite{mal}.

In realistic systems however, spin states remain constantly coupled to their environment.
Among various mechanisms of dephasing and decoherence of LZ transitions between Zeeman-split spin states, the coupling of a two-level system both with a phonon bath and a nuclear subsystem should be mentioned.
If the spin-nuclear coupling strength is weak enough and the relaxation of the nuclear bath is fast, then
the nuclear dynamics effects can be reduced to a fast random field \cite{kay, kay1984, kay1985, pok, pok2003}. In the opposite situation, if the nuclear subsystem is slow enough compared to the characteristic
tunnel time, its influence on LZ transition can be accounted by an effective model of
a slow classical noise\cite{Bergli, sai2001}.
For the simplest LZ scenario, the nuclear dynamics can be considered in the Born-Oppenheimer approximation, so that the spins are presumed isolated and transitions are activated by an external
magnetic field. Besides, the noise associated with both hyperfine and dipole fields plays an important role in the description of dynamical response of nanomagnets\cite{sai2001}.

Experiments with molecular magnets \cite{Sessoli1999}
revealed the presence of hysteresis phenomena in nanoscale molecular magnets among which are ${\rm Mn}_{12}$ and ${\rm Fe}_{8}$.
 LZ transitions at the avoided crossing between the Zeeman-split spin levels produced by hyperfine interactions have been pointed
out as responsible for plateaus on hysteresis loops\cite{Sessoli1999, kece}.
Hence, a number of proposals have been suggested, clarifying the effects of nuclear bath, noises and decoherence effects on the transition probability in
linearly driven systems including two- and multi-states systems\cite{ Volvo2004, Volvo2005, Volkov2007,  Sinitsyn, Sun2007, Sun2008, sai}.
Several compact analytic results have been derived to describe
these effects, namely, the Kayanuma's formula\cite{kay, kay1984, kay1985} for a strong diagonal noise and the Pokrovsky-Sinitsyn formula\cite{ pok, pok2003}
for the coupling to a fast colored noise with off-diagonal components.

Spin transport processes in magnetic semiconductor designs unavoidably suffer from hyperfine interactions treated
as a noise source frustrating spins during transmissions\cite{Loss, Fabian, Bett}.
The common way to protect information during the propagation consists on adiabatically applying an external
controlling magnetic field.
Betthausen et {\it al }\cite{Bett} have recently presented
an alternative experimental  method to protect spin propagation in spin transistors including diabatic LZ tunable
 transitions. Indeed, in these experiments,  a controlling magnetic field is a combination of a
spatially rotating magnetic field $B_{s}$ and a homogeneous field $B$. 
Thus, the spin states are subjected to both a constant magnetic field ($B$) and a fluctuating (Overhauser's) field.
A theoretical attempt to attack such a problem has been introduced in Refs.[\onlinecite{pok2003}], [\onlinecite{pok}] for two-level spin
systems by means of a fast noise associated with random hyperfine interactions.  In contrast to it, recent experiments
on the spin polarization of nuclear subsystem via time-dependent gate voltage in double
quantum dots\cite{Rashbba, Foletti2009, Foletti2010, Foletti2011} have shown that the fluctuations of the Overhauser's field are rather
slow, changing dramatically the properties of LZ transition. The "minimal theoretical model" should, however, take into
account all low-energy two-electron states in a dot consisting of three singlet and one triplet ($S=1$) states.

In this paper, we consider the influence of both fast and slow classical noise on two- and three- level systems.
We calculate transition probabilities for the noise-induced and noise-assisted processes
by using density matrix (Bloch) equation. The analytical expression for finite time probabilities
for two- and three- level systems are interpreted in terms of Fresnel's interference.
In addition to two standard definitions of a tunnel time for LZ transition by means of internal or external clocks,
we discuss a possibility to use noise as yet another probe for the LZ time.

The paper is organized as follows. Section \ref{Sec2} is devoted to the discussion of basic equations for LZ
transition $S=1/2$ derived through Schr\"odinger and Bloch approaches. In Sec.\ref{Sec3},
we discuss the noise-induced and noise-assisted LZ transitions in a two-level system. The classical noise
associated with fluctuations of the Overhauser's field is considered as a colored noise with the Gaussian realization.
Both the cases of one and two- component transverse noise are discussed. Sections \ref{Sec4} and \ref{Sec5} contain the key equations
for a three-level $S=1$ system subjected to both fast and slow classical noise. In Sec. \ref{Sec6}, we discuss the LZ transition times defined through internal and external clock in the presence of noise. The details of derivation
are sketched in Appendixes.

\section{Basic Relations for Two-level Systems}\label{Sec2}
\subsection{Schr\"odinger spin-$1/2$ picture}\label{Sec2.1}

The time evolution of $\mathcal{N}$ states of a quantum-mechanical system with a coherently driven total spin $S$  can be described by a system of $\mathcal{N}$ coupled differential equations for
the amplitudes $C_{1}^{(S)}(t),C_{2}^{(S)}(t),...,C_{\mathcal{N}}^{(S)}(t)$ of the states $\psi_{1}^{(S)}(t), \psi_{2}^{(S)}(t),...,\psi_{\mathcal{N}}^{(S)} (t)$ ($\hbar=1$):
\begin{align}\label{equ1}
i\dfrac{d}{dt}\mathbf{C}(t)=\hat{\mathcal{H}}(t)\mathbf{C}(t).
\end{align}
Here, $\mathbf{C}(t)=[C_{1}^{(S)},C_{2}^{(S)},...,C_{\mathcal{N}}^{(S)}]^{T}$ is a column vector for amplitude probabilities and
\begin{align}\label{equ2}
\hat{\mathcal{H}}(t)=\vec{\Theta}(t)\cdot\vec{S}
\end{align}
is the total Hamiltonian of the system, $\vec{S}$ is the total spin vector involving all the three generators of the group $SU(2)$.

The relevant aspect of Eq.(\ref{equ2}) with our aim lies in its description of Zeeman splitting of spin states in a linearly sweeping external magnetic field.
This aspect
   intimately refers  to the traditional LZ problem and the functions in Eq.(\ref{equ2}) are explicited as follows:
\begin{align}\label{equ3}
\Theta^{x}(t)=2\Delta, \quad \Theta^{y}(t)=0 \quad \textmd{and} \quad \Theta^{z}(t)=2\alpha t.
\end{align}
Here, $\alpha>0$ is the constant sweep velocity, $\Delta$ the tunneling coupling matrix element between states that we assume
here as real and varying from $t=-\infty$ to $t=\infty$.

For the special case of two levels, the problem (\ref{equ1}) leads to a system of two independent equations 
\begin{align}\label{equ3a}
\frac{d^{2}}{dz^{2}}C_{1}^{(1/2)}(z)+\Big[i\lambda-1/2-z^{2}/4\Big]C_{1}^{(1/2)}(z)=0,
\end{align}
\begin{align}\label{equ3b}
\frac{d^{2}}{dz^{2}}C_{2}^{(1/2)}(z)+\Big[i\lambda+1/2-z^{2}/4\Big]C_{2}^{(1/2)}(z)=0.
\end{align}
known as Weber's equations\cite{Erdelyi}, where $z=\sqrt{2\alpha}t e^{-i\pi/4}$ and $\lambda=\Delta^{2}/2\alpha$.
 Solutions of these equations are computed with respect to the initial conditions. For the choice $C_{1}^{(1/2)}(-\infty)=1$ and $C_{2}^{(1/2)}(-\infty)=0 $ i.e.
, when the particle was initially prepared in the state $\psi_{1}^{(1/2)}(t)$, one has\cite{lan,zen}:
\begin{align}\label{equ4}
C_{1}^{(1/2)}(t)=-\dfrac{A_{+}}{\sqrt{\lambda}} e^{-i\pi/4} e^{i\varphi}D_{-i\lambda}(-i\mu t),
\end{align}
and
\begin{align}\label{equ5}
C_{2}^{(1/2)}(t)=A_{-} e^{i\varphi}D_{-i\lambda-1}(-i\mu t).
\end{align}
Here, $D_{n}(z)$ is the parabolic  cylinder (Weber's\cite{Erdelyi}) function, $\varphi$ a phase factor and $\mu=\sqrt{2\alpha}e^{-i\pi/4}$. The parameter
$\lambda$ is introduced hereafter to distinguish between the sudden ($\lambda\ll1 $) and the adiabatic ($\lambda\gg1 $) limits of transitions.
The normalization factors $A_{+}$ and $A_{-}$ in Eqs.(\ref{equ4}) and (\ref{equ5}) are respectively defined by their modulus,
$\lvert A_{+}\rvert=\lvert A_{-}\rvert=\sqrt{\lambda}e^{-\pi\lambda/4}$.

The probability $\lvert C_{2}^{(1/2)}(t)\rvert^{2}$ that the system
will be found in the state $\psi_{2}^{(1/2)}(t)$ at any given time $t$ is therefore given by
\begin{align}\label{equ6}
P_{\textmd{LZ}}(t)=\lambda e^{-\pi\lambda/2}\lvert D_{-i\lambda-1}(-i\mu t) \rvert^{2}.
\end{align}
The symmetries of levels allow us to  directly find the probability to remain in the same state. Some asymptotic and exact values of Eq.(\ref{equ6}) are
 performed with the aid of an asymptotic series expansion of Weber's functions\cite{ Erdelyi}. By setting, for instance, $t\rightarrow\infty$, we recover
\begin{align}\label{equ7}
P_{\textmd{LZ}}(\infty)=1-e^{-2\pi\lambda},
\end{align}
known as the celebrated LZ formula\cite{lan, zen}. 

\subsection{Bloch spin-$1/2$ picture}\label{Sec2.2}

The general solution for the time-dependent LZ probability (\ref{equ6}) is written in terms of products
of Weber's functions. The asymptotic form of this equation casts, nevertheless, a very simple exponential form (\ref{equ7}). In this section, we present an approximate solution for LZ 
finite time probability (not necessarily in a long-time limit) as an exponential of a single-parametric real function and discuss the accuracy of this solution for sudden and adiabatic limits.

The time evolution of coherently driven quantum-dynamical system described by the model (\ref{equ2}) is here
governed by the von-Neumann equation for the total density matrix $\hat{\tilde{\rho}}(t)$,
\begin{align}\label{equ12}
i\dfrac{d\hat{\tilde{\rho}}(t)}{dt}=\Big[\hat{\mathcal{H}}(t),\hat{\tilde{\rho}}(t) \Big].
\end{align}
With the help of Eq.(\ref{equ12}), we find the population difference $\hat{\rho}(t)=\hat{\rho}_{11}(t)-\hat{\rho}_{22}(t)$ as being a solution
 of the differential equation
\begin{align}\label{equ13}
\dfrac{d\hat{\rho}(t)}{dt}=-i\Theta_{-}(t)\hat{\rho}_{21}(t)+i\Theta_{+}(t)\hat{\rho}_{12}(t),
\end{align}
where $\hat{\rho}_{21}(t)=\hat{\rho}_{12}^{*}(t)$ involving $\hat{\rho}^{*}(t)=\hat{\rho}(t)$ with,
\begin{align}\label{equ14}
\hat{\rho}_{12}(t)
=\dfrac{i\int_{-\infty}^{t}\exp\Big(i\int_{-\infty}^{t_{1}}\Theta^{z}(\tau')d\tau'\Big)\Theta_{-}(t_{1})\hat{\rho}(t_{1})dt_{1}}
{2\exp\Big(i\int_{-\infty}^{t}\Theta^{z}(\tau')d\tau'\Big)},
\end{align}
and $\Theta_{\pm}(t)=\Theta^{x}(t)\pm i\Theta^{y}(t)$. The indices 1 and 2 denote the two-level crossing. Inserting Eq.(\ref{equ14}) into Eq.(\ref{equ13}), with
reference to the context of LZ theory, i.e., $\Theta^{z}(t)=2\alpha t$, without loss of generality, we find the equation
\begin{align}\label{equ15}
\dfrac{d\hat{\rho}(t)}{dt}=-\int_{-\infty}^{t}dt_{1}\cos\Big[\alpha(t^{2}-t_{1}^{2})\Big]
\Theta_{+}(t)\Theta_{-}(t_{1})\hat{\rho}(t_{1}),
\end{align}
which can be included in the family of Volterra's integral-differential equations\cite{BurtonBook, Burton2005}. Equations (\ref{equ12})-(\ref{equ15}) correspond to Bloch's transformations related to the optical Bloch's\cite{Shore} equation
 $\dot{\vec{b}}=-\vec{\Theta}\times\vec{b}$, where $\vec{b}$ is the Bloch's vector set on a unit sphere
 by the condition $\textmd{Tr}\hat{\tilde{\rho}}= \textmd{Tr}\hat{\tilde{\rho}}^2 =1$. The $z$-component of it being expressed as a linear combination of diagonal matrix elements of $\hat{\tilde{\rho}}(t)$ as
$b_{z}(t)=\hat \rho(t)\equiv\hat{\rho}_{11}(t)-\hat{\rho}_{22}(t)$ satisfies Eq.(\ref{equ15}) 
\begin{align}\label{equ16}
\quad&&\hspace{-1.0205cm}\dfrac{d}{dt}\hat{\rho}^{(0)}(t)=-4\Delta^{2}\int_{-\infty}^{t}\cos\Big[\alpha(t^{2}-t_{1}^{2})\Big]\hat{\rho}^{(0)}(t_{1})dt_{1},
\end{align}
for the conventional LZ problem. The superscript $(0)$ refers to the LZ problem in the absence of noise.

The integral-differential equation (\ref{equ16}) can be solved iteratively with the condition $\hat{\rho}^{(0)}(-\infty)=1$ that preserves the total population at any arbitrary time $t$.
 A perturbation series expansion investigation with respect to the parameter $\lambda \ll 1$ (see Appendix.\ref{App1}) is achieved as,
\begin{align}\nonumber\label{equ17}
\hat{\rho}^{(0)}(t)=\nonumber1+2\Big[-2\pi\lambda F(t)+\dfrac{1}{2!}\Big(-2\pi\lambda F(t)\Big)^{2}+...\nonumber\\&&\hspace{-7cm}-G(t,\lambda)\Big]
=-1+2\Big[\exp\Big(-2\pi\lambda F(t)\Big)-G(t,\lambda)\Big].
\end{align}
The function $F(t)$ is defined as
\begin{align}\label{equ18}
F(t)=\dfrac{1}{2}\Big[\Big[c\Big(\sqrt{\dfrac{2\alpha}{\pi}}t\Big)+\dfrac{1}{2}\Big]^{2}+\Big[s\Big(\sqrt{\dfrac{2\alpha}{\pi}}t\Big)+\dfrac{1}{2}\Big]^{2}\Big].
\end{align}
$c(\sqrt{2\alpha/\pi}t)$ and $s(\sqrt{2\alpha/\pi}t)$ are respectively the cosine and sine Fresnel's integrals\cite{Abramowitz}.
The function $F(t)$ is more often encountered in the theory of light diffraction, where it relates to the intensity of light passed through a
 semi-infinite plane bounded by a sharp straight edge with $t$ assuming the lateral distance of the edge from the point of
observation\cite{Born}.

The function $G(t, \lambda)$ includes all corrections to the
exponential solution and is determined so that as $t\rightarrow\infty$ asymptotically one comes back to the
conventional LZ formula  (\ref{equ7}). Then, for convenience, we write our finite time transition probability as follows\cite{Kiselev}:
\begin{align}\label{equ19}
P_{\textmd{LZ}}(t)=1-\exp\Big\{-2\pi\lambda[F(t)+\ln W(t)]\Big\},
\end{align}
with
\begin{align}\label{equ20}
\quad&&\hspace{-.93cm}\ln W(t)=-\dfrac{1}{2\pi \lambda}\ln\Big[1-G(t, \lambda)
\exp\Big\{2\pi\lambda F(t)\Big\}\Big],
\end{align}
in which
\begin{align}\label{equ21}
G(t,\lambda)=\lambda e^{-\pi\lambda/2}\lvert D_{-i\lambda-1}(-iz)\rvert^{2}-\Big(1-e^{-2\pi\lambda F(t)}\Big).
\end{align}
One can see that in the limit $\lambda \ll 1$  the correction function $\ln W(t) < F(t)$ for all times.
The two forms Eq.(\ref{equ19}) and Eq.(\ref{equ6}) are equivalent with
 the only difference being that Eq.(\ref{equ19}) is the exponential form of Eq.(\ref{equ6}).
The form (\ref{equ19}) we obtained will be used for analytic derivations of finite-time transition probabilities in the limit of slow noise driven LZ transitions.

The Fresnel's integrals give rise to Fresnel's oscillations (see Fig.\ref{FIG1} below) and suggest interferences between states around the
anticrossing region. The Fresnel-type oscillatory factors $e^{\pm i\alpha t^{2}}$ involved in Fresnel's integrals originate from the phase
\begin{align}\label{equ21'}
\int_{0}^{t}\Theta^{z}(t') dt'=\alpha t^{2},
\end{align}
accumulated by the two components of the wave function during a linear sweep. 

\begin{figure}[]
 \begin{center}
    \leavevmode
    \subfloat[]{%
      \label{fig1a}
      \includegraphics[width=4.3cm, height=35mm]{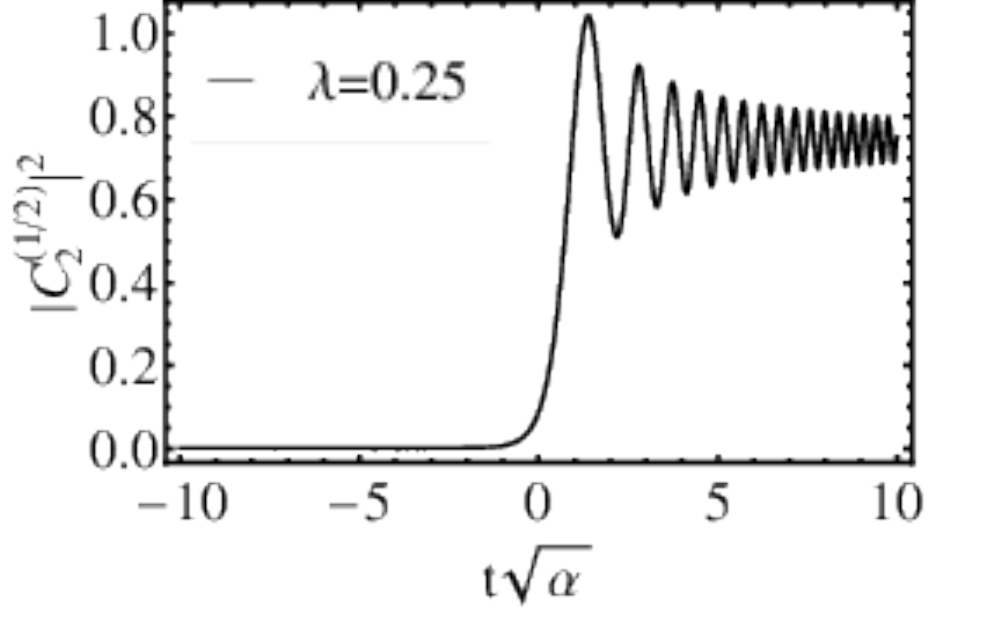}}
    \hspace{-6mm}
    \subfloat[]{%
      \label{fig1b}
      \includegraphics[width=4.3cm, height=35mm]{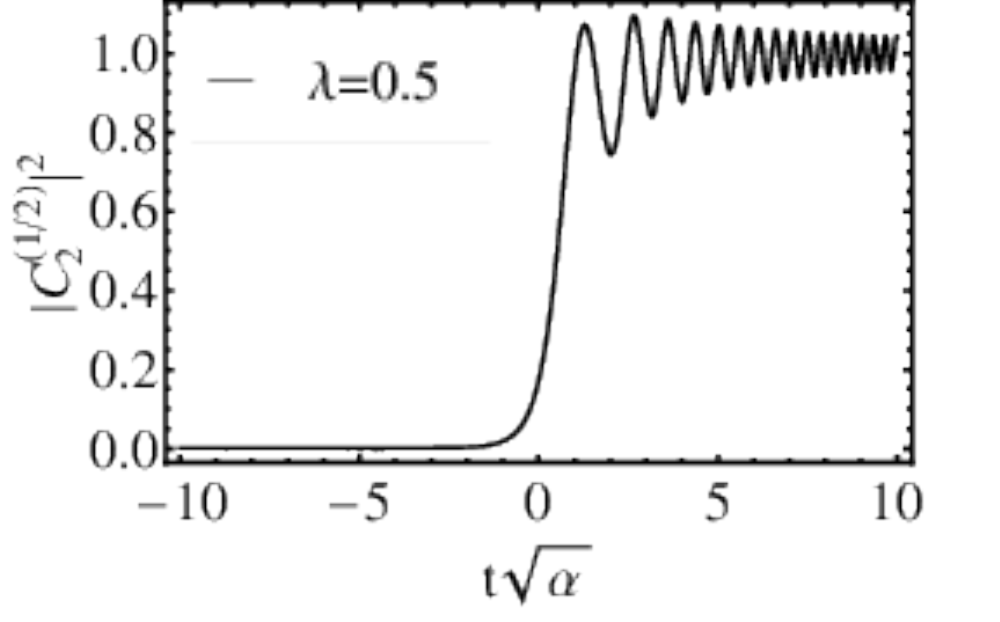}}
    \hspace{-6mm}
\subfloat[]{%
      \label{fig1c}
      \includegraphics[width=4.3cm, height=35mm]{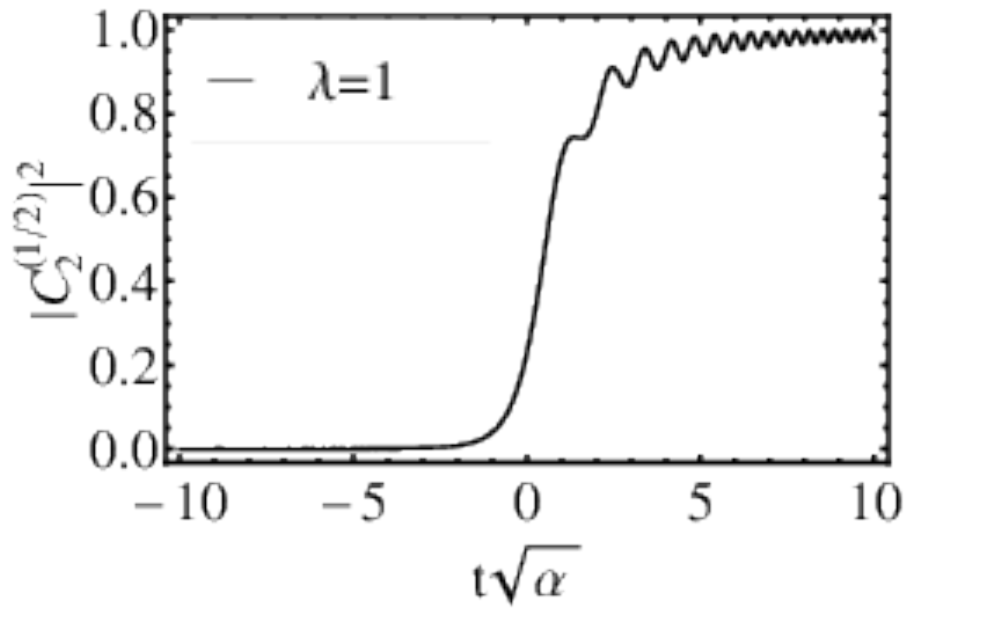}}
    \hspace{-6mm}
\subfloat[]{%
      \label{fig1d}
      \includegraphics[width=4.3cm, height=35mm]{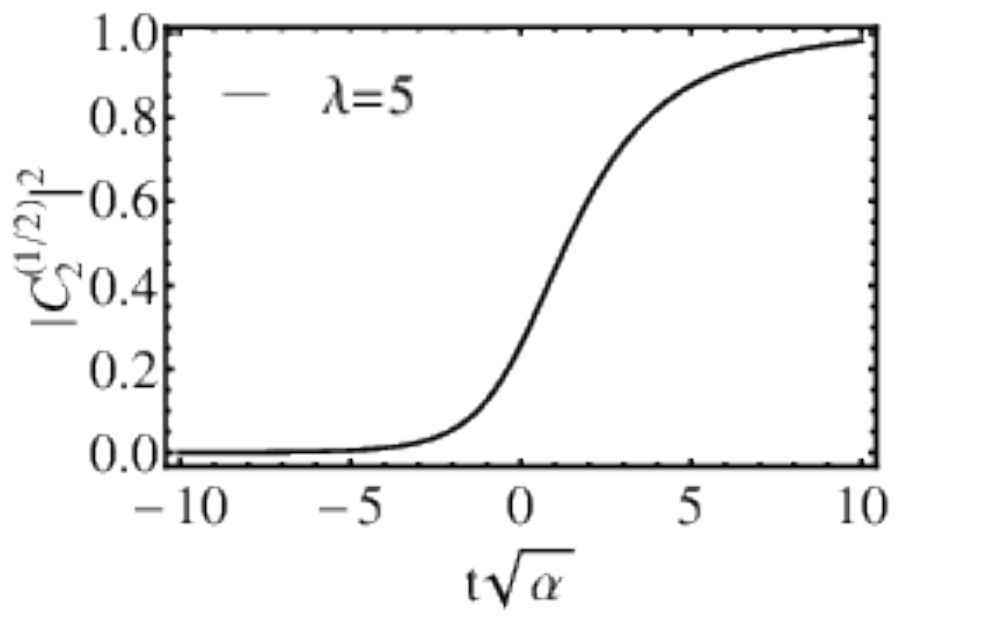}}
 \end{center}
\caption{Time evolution of the LZ transition probability in the diabatic basis of a two-level system.
(a) The small values of the Landau-Zener parameter $\lambda=\Delta^2/2\alpha$ correspond to sudden limit (rapid drive). (d) 
The large values of $\lambda$ describe adiabatic limit (slow drive).  The intermediate LZ regimes are presented in (b) and (c). The time is in units of $1/\sqrt{\alpha}$.}\label{FIG1}
\end{figure}

The function $P_{\textmd{LZ}}(t)$ could have also  been found by numerically solving the time-dependent Schr\"odinger's equation (\ref{equ1}).
A typical result is depicted by Fig.\ref{FIG1} above, where the time evolution of the transition probability $P_{\textmd{LZ}}(t)$ discriminates two regions. In the domain $t\leq0$, all probabilities are smooth monotonic functions of time.
First, around the crossing time $t=0$, one sees a sharp transition of $P_{\textmd{LZ}}(t)$ which rises from zero to its maximum value and later
behaves asymptotically with decaying oscillations around the saturation value $P_{\textmd{LZ}}(\infty)$. This last performance characterizes the system for small values of $\lambda$ (sudden limit),
while for largest $\lambda$ (adiabatic limit), oscillations are strongly mitigated.
The first region identified characterizes the jump time $\tau_{\textmd{jump}}$ while the second determines the relaxation time $\tau_{\textmd{relax}}$\cite{Vitanov1999}. The jump time $\tau_{\textmd{jump}}$
called hereafter LZ transition time is denoted as $\tau_{\textmd{LZ}}$. 
In adiabatic limit (slow passage), this time should be a function of the velocity $\alpha$  and the splitting energy $\Delta$ (see discussion in Sec.\ref{Sec6}):
\begin{align}\label{equation22}
\tau_{\textmd{LZ}}=\mathcal{L}(\alpha, \Delta).
\end{align}

The rapid drive of a two-level system produces repeated LZ transitions after passing the resonance, drives the system into
a coherent superposition of states where they may interfere and generates St\"uckelberg's fringes\cite{Oliver, Sillan, Wilson, Mark, Levitov2006, Levitov2008}.
As a consequence, appearing fringes encode information about the system's evolution and energy spectrum\cite{stu, Oliver, Sillan, Wilson, Mark, Levitov2006, Levitov2008}.
 The system will not then feel the gap and $\tau_{\textmd{LZ}}$ should be independent on the energy splitting $\Delta$:
\begin{align}\label{equation23}
\tau_{\textmd{LZ}}=\mathcal{L}(\alpha).
\end{align}
The slow drive in the opposite extreme limit produces oscillations of very weak amplitudes so that one can assume a single transition; St\"uckelberg's fringes of the former sort
 could be absent on the interferometer. The advantage being the possibility to probe spectroscopic information about the coherent evolution of the system\cite{Levitov2008}. As the system feels the gap,
\begin{align}\label{equation24}
 \tau_{\textmd{LZ}}=\mathcal{L}(\alpha, \Delta).
\end{align}
Semiclassically, $\tau_{\textmd{LZ}}$ is the time necessary to reach a turning point on the imaginary time axis of the integration  contour\cite{QM}.

This mechanism of slowing down the sweep rate in order to collect spectroscopic information about a quantum system was recently employed by Berns and co-workers in their experiment of
spectroscopy analysis of a solid-state artificial atom\cite{Levitov2008}. In that experiment, they pointed out that for St\"uckelberg's interference to
occur, the time interval between two consecutive LZ transitions should be much more smaller than the relevant decoherence times.

 An alternative way to tackle the traditional LZ problem consists on transforming Eq.(\ref{equ16})
 to a differential equation. It can be achieved by applying a second-order time derivative to both sides of equation (\ref{equ16}) and excluding the integral term with $\sin\left[\alpha(t^2-t_1^2\right]$.
 Hence, we show that $\hat{\rho}^{(0)}(t)$ satisfies the third-order differential equation,
\begin{align}\nonumber\label{equ25}
&&\hspace{-3cm}\dfrac{d^{3}}{d\tau^{3}}\hat{\rho}^{(0)}(\tau)-\dfrac{1}{\tau}\dfrac{d^{2}}{d\tau^{2}}\hat{\rho}^{(0)}(\tau)-4\Big[\dfrac{2\lambda}{\tau}\hat{\rho}^{(0)}(\tau)\\&&\hspace{2cm}-(\tau^{2}+2\lambda)\dfrac{d}{d\tau}\hat{\rho}^{(0)}(\tau)\Big]=0,
\end{align}
which can be interpreted as a differential equation for the probability. Here, we performed the time scaled transformation $\tau=t\sqrt{\alpha}$. While amplitudes found from
the linear Schr\"odinger's equation satisfy a second-order linear differential equation, probabilities from the von-Neumann equation rather satisfy a third order linear differential equation.
 A similar equation was written in Refs. \onlinecite{Nakamura2005, Nakamura2006, Izmalkov, Vitanov1999} in the same context of LZ theory and numerically solved in Ref.\onlinecite{Vitanov1999} with the aid of a Runge-Kutta algorithm.
Here, using a correspondence between Schr\"odinger and Bloch approaches we propose an analytic and exact solution to this kind of equations.

The natural initial condition $\hat{\rho}^{(0)}(-\infty)=1$ was gradually translated at each step of derivations and we solve Eq.(\ref{equ25}) with the
 conditions
\begin{align}\label{equ26}
\dfrac{d^{2}}{d\tau^{2}}\hat{\rho}^{(0)}(\tau)\mid_{\tau=-\infty}=-8\lambda, \quad \dfrac{d}{d\tau}\hat{\rho}^{(0)}(\tau)\mid_{\tau=-\infty}=0.
\end{align}
As $\hat{\mathcal{H}}(t)$ in the Schr\"odinger's equation (\ref{equ1}) realizes $SU(2)$ symmetry operations, the amplitudes $C_{1}^{(1/2)}(\tau)$  and
$C_{2}^{(1/2)}(\tau)$ also realize the same set of operations. This is due to the temporal linearity of the Schr\"odinger's equation
which preserves symmetry properties. 
While a Schr\"odinger's equation describes dynamics of wave functions [$SU(2)$ spinors], the Bloch's equation deals with the evolution of probability densities combined into a vector on a unit sphere (Bloch's vector on 2-sphere). However, a local isomorphism  between $SU(2)$ and $SO(3)$ establishes relations between these two objects.

 In the population difference, the occupation probabilities are expressed in terms of transition amplitudes for half-spin
in Eqs.(\ref{equ4}) and (\ref{equ5}) i.e.: $\hat{\rho}_{11}^{(0)}(\tau)=\lvert C_{1}^{(1/2)}(\tau)\rvert^{2}$ and $\hat{\rho}_{22}^{(0)}(\tau)=\lvert C_{2}^{(1/2)}(\tau)\rvert^{2}$ and this realizes the isomorphism we talked about.
Thus, the solution of Eq.(\ref{equ25}) reads,
\begin{align}\nonumber\label{equ27}
\hat{\rho}^{(0)}(\tau)=-\lambda e^{-\pi \lambda/2}\Big[\lvert D_{-i\lambda-1}(-i\mu_{0} \tau)\rvert^{2}\\&&\hspace{-2.2cm} -\dfrac{1}{\lambda}\lvert D_{-i\lambda}(-i\mu_{0} \tau)\rvert^{2} \Big].
\end{align}
Here, $\mu_{0}=\mu/\sqrt{\alpha}$. 
We may deduce, from the same technique, an integral relation between Weber's functions. From Eq.(\ref{equ14}) one may notice that
$\hat{\rho}_{12}^{(0)}(t)=C_{1}^{(1/2)}(t)C_{2}^{(1/2)*}(t)$ and find
\begin{widetext}

\begin{align}\label{equ28}
 D_{-i\lambda }(-i\mu t)D^{*}_{-i\lambda-1}(-i\mu t)=
-\lambda\mu\int^{t}_{-\infty }\exp\Big(\dfrac{\mu^{2}}{2}[t^2-t_{1}^2]\Big)\Big(\lvert D_{-i\lambda-1}(-i\mu t_{1})\rvert^{2}- \dfrac{1}{\lambda}\lvert D_{-i\lambda}(-i\mu t_{1})\rvert^{2} \Big)dt_{1}.
\end{align}
\end{widetext}

A similar relation for $D_{-i\lambda }^{*}(-i\mu t)D_{-i\lambda-1}(-i\mu t)$ can be derived from $\hat{\rho}_{21}^{(0)}(t)=C_{1}^{(1/2)*}(t)C_{2}^{(1/2)}(t)$.
A similar matching procedure was recently employed in Ref.[\onlinecite{Rashbba}] to establish an integral relation between Weber's functions
 not from $\hat{\rho}_{12}^{(0)}(t)$ as we did here but from $\hat{\rho}^{(0)}(t)$  in Eq.(\ref{equ13}).

One can easily check that the limit $\tau \to \infty$ applied to Eq.(\ref{equ27}) brings us back automatically to Eq.(\ref{equ7}).
Thus, the former represents the population difference at any given time $\tau$.
The solution of Eq.(\ref{equ25}) gives information about the time dependence of population difference 
directly measurable in the flux qubits experiments in a micromaser \cite{G.Sun2009, Rouse1996, You2007}.
 It might serve for transfer of population between two states at any time $\tau$. For instance by measuring the LZ transition probability between two states,
 it provides information about the strength $\Delta$ of the coupling between states.
It could also offer great advantages in experiments with atoms transfer, having only one parameter for control.

In the domain $\tau\leq0$, the projection of Bloch's vector on $z$-axis is positive.
The system remains in the state where it has been set initially. Passing
now through the resonance, $\hat{\rho}^{(0)}(\tau)$ abruptly changes its concavity becoming either greater or less than zero. Just around the anticrossing region, the sharp drop
of $\hat{\rho}^{(0)}(\tau)$ shows that $\lvert 1 \rangle$ has started to feed $\lvert 2 \rangle$ via the LZ mechanism.

 In the domain $\tau>0$, one has $\rho^{(0)}(\tau)<0$, the two-level system experiences decaying oscillations
 while the population difference saturates to a finite value. The oscillations correspond to an interference between states $\lvert 1 \rangle$ and $\lvert 2 \rangle$.
This last remark tells us that population difference tends to maintain the majority
of the system into the excited state rather than the ground state. 

\section{Transverse Noise in the spin-$1/2$ Landau-Zener theory}\label{Sec3}

We now turn into a situation where LZ transitions are noise induced. Basically, the coupling between level positions fluctuates due to a transverse noise with the Gaussian realizations.

LZ effects in the presence of transverse classical noise including inter-level transitions are specified by the prototype Hamiltonian (\ref{equ2}) considering
\begin{align}\label{equ22}
\Theta^{x}(t)=2f_{x}(t), \quad \Theta^{y}(t)=2f_{y}(t) \quad \textmd{and} \quad \Theta^{z}(t)=2\alpha t.
\end{align}
These definitions are also valid for the case $S=1$ we study below. The mean-zero stochastic functions $f_{i}(t)$($i=x,y$) in Eq.(\ref{equ22}) 
are characterized by their first- and second- order moments,
\begin{align}\label{equ23}
\langle f_{i}(t)\rangle=0, \quad \langle f_{i}(t)f_{j}(t')\rangle=\eta^{2}\delta_{ij}\exp({-\gamma\lvert t-t' \rvert}).
\end{align}
Here, $\eta$ stands for the noise intensity that might be related to the absolute temperature via the universal {\em fluctuation dissipation theorem}\cite{Havas} (see detailed discussion below). The parameter
$\gamma=1/t_{noise}$ defines a time scale associated with the noise. Comparison of $t_{noise}$ with characteristic time scales of LZ problem gives us a definition of fast and slow noise limits.
The dynamics of the system is governed by Eq.(\ref{equ15}) for the occupation difference:
\begin{align}\label{equ24}
\quad&&\hspace{-1.0205cm}\dfrac{d\hat{\rho}(t)}{dt}=-4\int_{-\infty}^{t}\cos\Big[\alpha (t^{2}-t_{1}^{2})\Big]f_{+}(t)f_{-}(t_{1})\hat{\rho}(t_{1})dt_{1},
\end{align}
where $f_{\pm}(t)=f_{x}(t)\pm if_{y}(t)$. The solution of this equation is to be averaged over all possible realizations of the two-level system (ensemble average).
The result of this averaging is different for the two limits of fast and slow noise.

\subsection{Fast noise, spin-$1/2$}\label{Sec3.1}

If a noise is fast, the characteristic noise time
 $t_{noise}\ll \tau_{\textmd{LZ}}$, one can average Eq.(\ref{equ24}) directly and decouple the product
 $\langle f_{+}(t)f_{-}(t_{1})\hat{\rho}(t_{1})\rangle$ as $\langle f_{+}(t)f_{-}(t_{1})\rangle\langle\hat{\rho}(t_{1})\rangle$. The resulting master equation
 for the average $\langle\rho(t)\rangle$ gives the conventional equation for the transition probability as the average $\langle\hat{\rho}(t_{1})\rangle$
 does not really change in the exceedingly short time interval $\xi=t_{1}-t$:
\begin{align}\label{equa29}
\dfrac{d}{dt}\langle\hat{\rho}(t)\rangle=-\hat{\Omega}(t)\langle\hat{\rho}(t)\rangle.
\end{align}
Here, the functional $\hat{\Omega}(t)\equiv\hat{\Omega}[\tilde{\omega}(t)]$ of the frequency $\tilde{\omega}(t)=2\alpha t$ is defined through the two-time correlation function $\hat{\mathcal{R}}(\lvert t-t_{1}\rvert)=\langle f_{+}(t)f_{-}(t_{1})\rangle$ as
$\hat{\Omega}(t)=\hat{\Omega}^{(+)}(t)+\hat{\Omega}^{(-)}(t)$, where
\begin{align}\label{equa30}
\hat{\Omega}^{(\pm)}(t)=\int_{-\infty}^{\infty}\exp\Big[\pm i\tilde{\omega}(t) \xi\Big]\hat{\mathcal{R}}(\lvert \xi \rvert)d\xi,
\end{align}
are the power spectral densities of noise capturing information about environmental effects.
For the Gaussian model we considered, Eq.(\ref{equa30}) result in a
Lorentzian. The frequency $\tilde{\omega}(t)$ is antisymmetric $\tilde{\omega}(-t)=-\tilde{\omega}(t)$,
 the Lorentzian spectral density is symmetric in the Fourier space leading thus to $\hat{\Omega}^{(+)}(t)=\hat{\Omega}^{(-)}(t)$.
Equation (\ref{equa29}) is thus  readily solved to give:
\begin{align}\label{equa31}
\langle\hat{\rho}(t)\rangle=\langle\hat{\rho}(-\infty)\rangle\exp\Big[-\int_{-\infty}^{t}\hat{\Omega}(t')dt'\Big],
\end{align}
and $\hat{\Omega}(t)$ is readily integrated accounting for the spectral density. As a result,
 we obtained the phase $\vartheta(t)$ accumulated during an interval of time $(-\infty,t]$:
\begin{align}\label{equa31a}
\vartheta(t)=\int_{-\infty}^{t}\hat{\Omega}(t')dt'=\frac{\pi\hat{\mathcal{R}}(0)}{\alpha}\Big[1+\dfrac{2}{\pi} \arctan(\dfrac{2\alpha}{\gamma}t)\Big].
\end{align}
and $\theta(t)=2\vartheta(t)$. By putting as usual $t=\infty$ one gets
\begin{align}\label{equa32}
\theta(\infty)=2\int_{-\infty}^{\infty}\hat{\Omega}(t')dt'=\dfrac{4\pi}{\alpha}\hat{\mathcal{R}}(0).
\end{align}
So, the transition probability results from Eqs.(\ref{equa31}) and (\ref{equa31a}) or (\ref{equa32}) as a combination of the initial condition $\hat{\rho}(-\infty)=1$ and
 the conservation law $\textmd{Tr}\hat{\tilde{\rho}}(t)=1$:
\begin{align}\label{equa38}
P_{fn}=\frac{1}{2}\Big[1-e^{-\theta/2}\Big].
\end{align}
Here, $\theta=\theta(\infty)$. Equation (3.9) is generalized  by Pokrovsky\cite{pok2003, pok}  to arbitrary correlation function.   It demonstrates an equal distribution of the system between
the ground and excited states after passing the crossing time for large noise $\eta\rightarrow\infty$. By taking the limits $\eta\to\infty$ and $\gamma\to\infty$ while keeping $\eta^{2}/\gamma=const$, the white noise
limit can be obtained from (\ref{equ23}). Note, that $\theta\sim \eta^2$ does not depend on $\gamma$ in that limit.

\begin{figure}[!h]
  \begin{center}
    \leavevmode
    \subfloat[]{%
      \label{fig3a}
      \includegraphics[width=4.3cm, height=35mm]{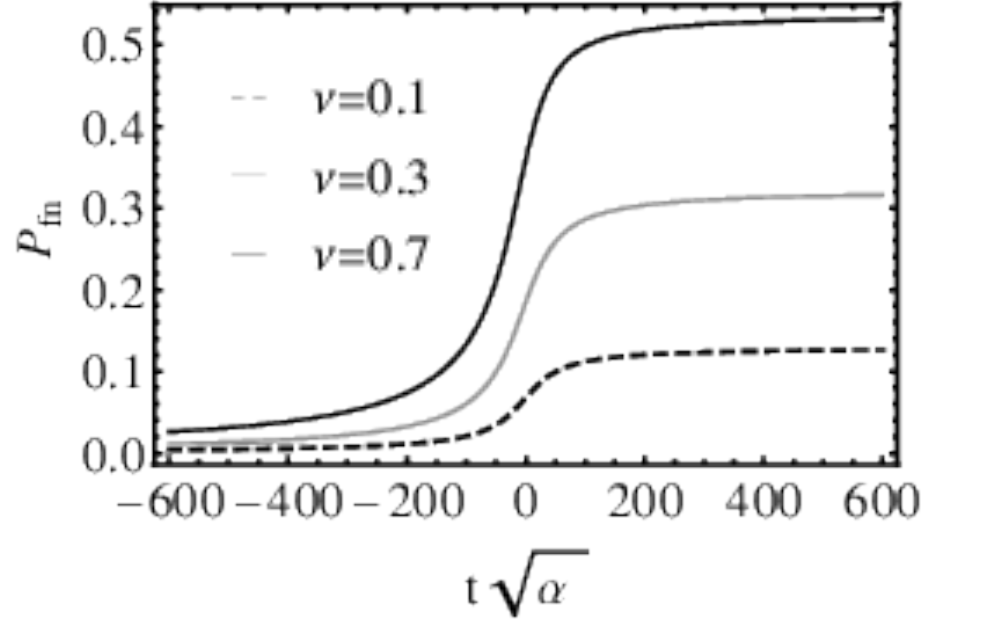}}
    \hspace{-4mm}
    \subfloat[]{%
      \label{fig3b}
      \includegraphics[width=4.3cm, height=35mm]{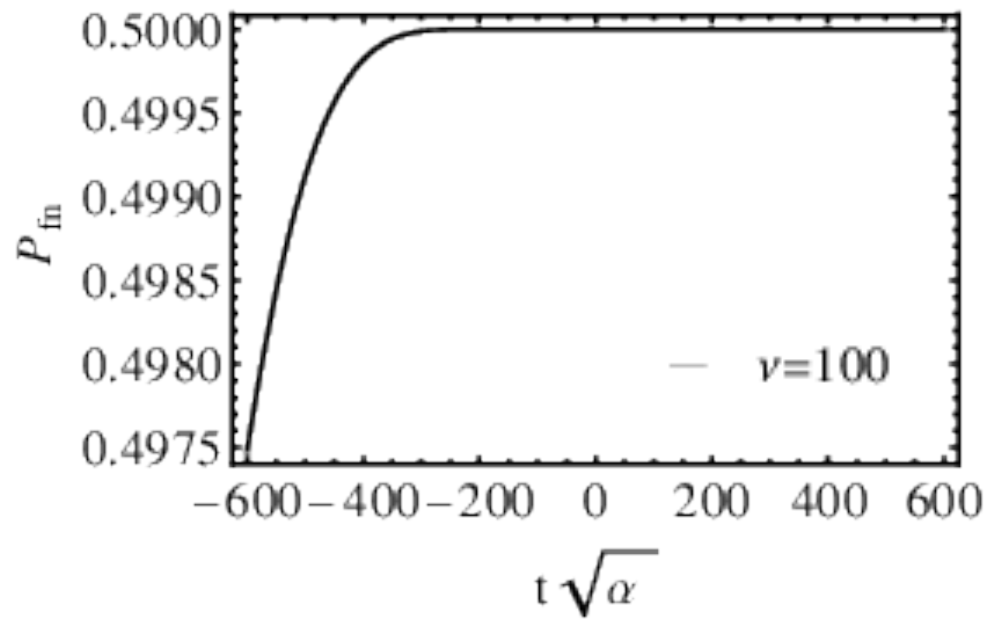}}
 \end{center}
\caption{ Time evolution of the LZ transition probability in the diabatic basis of a two-level system in presence of a fast
transverse noise for the rapid (a) and slow (b) passages. The amplitude of the fast noise is fixed. The noise is characterized by a dimensionless parameter $\nu=\pi\eta^2/\alpha$ and dimensionless frequency $\gamma_0=\gamma/\sqrt{\alpha}$. For all calculations $\gamma_{0}=100$.
} \label{FIG3}
\end{figure}

Here, we defined the dimensionless frequency $\gamma_{0}=\gamma/\sqrt{\alpha}$ and dimensionless parameter $\nu=\pi\eta^2/\alpha$.
If a noise is directed on either the transverse direction ($X$ noise) or the two-components transverse noise
 ($XY$ noise), Eq.(\ref{equa38}) reduces respectively to
\begin{align}\label{equa39}
P_{fn}^{x}=\frac{1}{2}\Big[1-\exp\Big(-\frac{2\pi}{\alpha}\langle f_{x}(t)f_{x}(t)\rangle\Big)\Big]
\end{align}
and
\begin{align}\label{equa40}
P_{fn}^{xy}=\frac{1}{2}\Big[1-\exp\Big(-\frac{2\pi}{\alpha}\Big[\langle f_{x}(t)f_{x}(t)\rangle+\langle f_{y}(t)f_{y}(t)\rangle\Big]\Big)\Big].
\end{align}
Hence, in order to sum up noises in $X$ and $Y$-directions it just suffices to do that in the argument of the exponential in Eq.(\ref{equa38}).
What happens if the noise is colored in one direction and white in another? The answer to this question is provided by the argument of the
exponential in Eq.(\ref{equa40}). Obviously, white noise will dominate the colored one and there will not be a complete transfer of population:
 both states remaining constantly coupled.

The solution $(\ref{equa31a})$ generalizing Eq.$(\ref{equa32})$ to finite times coincides at very large times with the results obtained in Refs.\onlinecite{pok, pok2003}. Similar results were discussed in Ref.\onlinecite{Bergli}
 for nonlinear drive with telegraph noise. A relevant note similar to the fast telegraph noise in
 a two-level system for this Gaussian model is that the noise fluctuations are averaged out as there are no fluctuations (see Fig. \ref{FIG3}).

We specify for further purposes that $\eta^{2}/\alpha\ll1$ and $\eta^{2}/\alpha\gg1$  correspond, respectively, hereafter to the sudden and
 adiabatic limits of transitions.
In the adiabatic limit then, the transition probability depends non-analytically on the sweep velocity.
Thus far, for $t_{noise}\ll \tau_{\textmd{LZ}}$, there
 is no complete transfer of population; the two states are constantly occupied.

\subsubsection*{Spin-$1/2$, in a constant off-diagonal field and a fast transverse random field}\label{Sec3.1.1}

Let us consider a spin coupled to a constant off-diagonal and sweeping magnetic field and a transverse noise source.
Such pattern corresponds for instance to spin frustrated by a hyperfine field or the Overhauser's field and protected by a constant magnetic field.
With reference to the recent experimental work in Ref.\onlinecite{Bett}, we present an alternative way to protect
spin propagation in spin transistor (see Introduction). In this frame, LZ transitions are noise assisted and noise fields are no longer centered at the origin in the $X$-direction:
\begin{align}\label{equa41}
f_{x}(t)=\Delta+\tilde{f}_{x}(t), \quad f_{y}(t)=\tilde{f}_{y}(t).
\end{align}
Noise correlation functions for $\tilde{f}_{i}(t)$ are given by Eq.(\ref{equ23}). The model (\ref{equ22}) with Eq.(\ref{equa41}) can also assume a
spin weakly interacting with an environment, for example a nuclear spin bath.  Assuming the spin-bath interaction as weak enough as bath relaxation is
much faster than the inverse interaction energy, we may treat the bath as a fast noisy magnetic field\cite{pok}.

 As far as noise is no longer centered at the origin, we call $\hat{\rho}^{(\textmd{SF})}(t)$ the average of the
total density matrix for the non-zero-mean problem labeled by Eq.(\ref{equa41}).
 Straightforward calculations for spin-$1/2$ suggest a governing equation of the form
\begin{align}\nonumber\label{equa42}
\dfrac{d}{dt}\hat{\rho}^{(\textmd{SF})}(t)=-4\Delta^{2}\int_{-\infty}^{t}dt_{1}\cos[\alpha(t^{2}-t_{1}^{2})]
\hat{\rho}^{(\textmd{SF})}(t_{1})&&\hspace{-2cm}\\-4\int_{-\infty}^{t}dt_{1}\cos[\alpha(t^{2}-t_{1}^{2})]\tilde{f}_{+}(t)\tilde{f}_{-}(t_{1})
\hat{\rho}^{(\textmd{SF})}(t_{1}).
\end{align}
Linear terms of noise have been dropped since after averaging and use of fast noise requirements they vanish.
Hereafter, we adopt the label $\mathcal{P}$ to denote transition
probabilities related to the non-centered noise.

As noise is fast, we can readily average Eq.(\ref{equa42}) as we did before and apply the decoupling procedure associated with the
 other arguments of fast noise.  As a result, the differential equation casts a form
\begin{align}\nonumber\label{equa43}
&&\hspace{-.2cm}\dfrac{d}{dt}\langle\hat{\rho}^{(\textmd{SF})}(t)\rangle=-4\Delta^{2}\int_{-\infty}^{t}dt_{1}\cos[\alpha(t^{2}-t_{1}^{2})]
&&\hspace{-.2cm}\langle\hat{\rho}^{(\textmd{SF})}(t_{1})\rangle\\&&\hspace{-7cm}-\hat{\Omega}(t)\langle\hat{\rho}^{(\textmd{SF})}(t)\rangle.
\end{align}

We obtained the conventional master equation for the transition probability in which noise appears as a perturbing source. Noise
essentially modifies the standard occupation difference $\hat{\rho}^{(0)}(t)$ by a decaying random phase factor. Mainly, noise produces
 dephasing during the transfer of population. The phase accumulated during the short time interval $t_{1}-t\sim1/\gamma$ is small enough such that:
\begin{align}\label{equa44}
\int_{-\infty}^{t_{1}}\Omega(t')dt'\approx \int_{-\infty}^{t}\Omega(t')dt'.
\end{align}
Indeed, as the characteristic frequency $\gamma \to \infty$ it appears that $t_{1}\approx t$, justifying the approximation Eq.(\ref{equa44}).
Thus, the solution of (\ref{equa43}) can be formally expressed as follows:
\begin{align}\label{equa45}
\langle\hat{\rho}^{(\textmd{SF})}(t)\rangle=\exp\Big[-\int_{-\infty}^{t}\hat{\Omega}(t')dt'\Big]\hat{\rho}^{(0)}(t).
\end{align}

For the solution at $t=\infty$, we derive $\hat{\rho}^{(0)}(\infty)=2e^{-2\pi\lambda}-1$ from Eq.(\ref{equ27}) and the finding transition probability
is obtained as follows:
\begin{align}\label{equa46}
\mathcal{P}_{fn}^{(\textmd{SF})}=\frac{1}{2}\Big[1-e^{-\theta/2}(2e^{-2\pi\lambda}-1)\Big].
\end{align}
Equation.(\ref{equa46}) describing the probability to remain in the same adiabatic state is fully consistent with the one obtained in Refs. \onlinecite{pok}. Namely,
 Eq.(42) in Ref.  \onlinecite{pok}  describes a system initially set in the diabatic state $\rvert1\rangle$ and conserves the same state while Eq.(\ref{equa46}), describes a spin flip between two different diabatic states. 

As it is discussed in Ref. \onlinecite{pok2003}, the fast noise can lead to full equilibration depending on the time-scales involved (see Sec. VI of Ref. \onlinecite{pok2003} for detailed analysis).  In Eq.(\ref{equa46}), assuming an adiabatic addition of noise $\theta \to \infty$ ($\alpha \to 0$), the probability achieves the value $1/2$. Such a system loses its memory.
It becomes obvious that by setting $\theta=0$ we recover the LZ formula (\ref{equ7}). Likewise, the requirement $\lambda=0$
leads to the solution for fast noise centered at origin.  Spin state evolution in spin-transistors might be protected during the transport by adiabatically applying a homogeneous magnetic field ($\lambda\gg1$).
 This technique was employed in Ref.[\onlinecite{Bett}].

\begin{figure}[t]
\begin{center}
 \includegraphics[width=7cm, height=4.5cm]{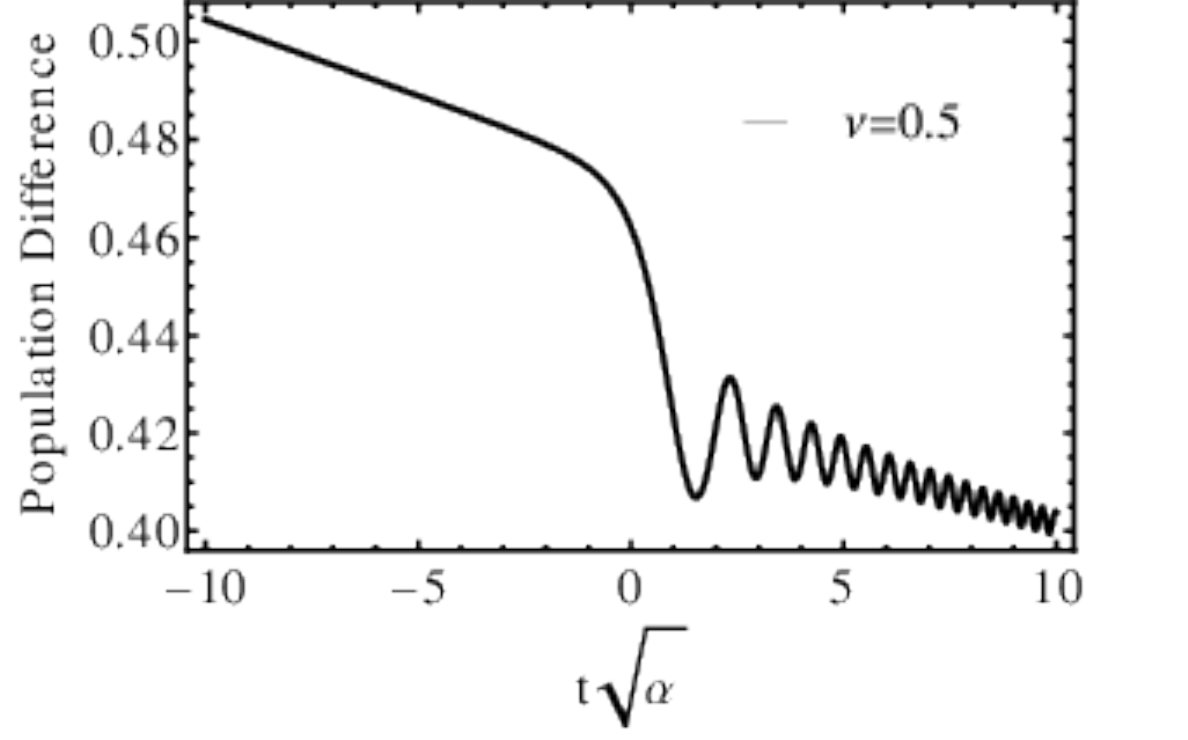}
\end{center}
\caption{Dynamical evolution of the population difference $\langle\hat{\rho}^{(\textmd{SF})}(t)\rangle$ Eq.(\ref{equa43}) in the presence of a fast transverse noise with characteristic decay rate $\gamma_0=\gamma/\sqrt{\alpha}$ and amplitude $\nu=\pi\eta^2/\alpha$ for a rapid LZ drive $\lambda=\Delta^2/2\alpha\ll 1$. The numerical calculations are performed with the parameters 
$\gamma_{0}=100$, $\nu=0.5$ and $\lambda=0.05$.} \label{FIG4}
\end{figure}

In Fig. \ref{FIG4}, the abrupt decay of the population difference $\langle\hat{\rho}^{(\textmd{SF})}(t)\rangle$ around the anti-crossing region characterizes a
rapid transfer of population. However, as the condition $\langle\hat{\rho}^{(\textmd{SF})}(t)\rangle>0$
 is always fulfilled, there is no way to expect a complete transfer from one of the diabatic states to another with the fast noise. Fast noise being characterized by short time memory,
 essentially creates a dephasing between the states of a two-level system.

\subsection{Slow noise, spin-$1/2$}\label{Sec3.2}

If now noise is slow ($t_{noise} \gg \tau_{\textmd{LZ}}$), the decoupling procedure is not applicable, thus the density matrix equation of motion (\ref{equ24}) cannot be
reduced to a master equation. Instead one has to solve Eq.(\ref{equ24}) and perform ensemble average over the distribution
$Q$ of noise:
\begin{align}\label{equa47}
\langle...\rangle=\dfrac{1}{\sqrt{2\pi}\eta}\int_{-\infty}^{\infty}dQ...\exp\Big(-\dfrac{Q^{2}}{2\eta^{2}}\Big).
\end{align}
The brackets $\langle...\rangle$ indicate as usual the ensemble average. In a given realization of classical field $Q$, 
the LZ probability is given by the standard equation
\begin{align}\label{equ31}
P_{\textmd{LZ}}(Q)=1-\exp\Big(-\dfrac{\pi Q^{2}}{\alpha}\Big).
\end{align}
If the noise in transverse direction is single-component, we can always rotate our coordinate frame such a
way that the fluctuations occur along $X$-direction. The noise - averaged LZ probability  is defined as 
\begin{align}\label{equ32}
P_{sn}^{x}=\dfrac{1}{\sqrt{2\pi}\eta}\int_{-\infty}^{\infty}dxP_{\textmd{LZ}}(x)\exp\Big(-\dfrac{x^{2}}{2\eta^{2}}\Big),
\end{align}
and after straightforward calculation is given by
\begin{align}\label{equ33}
P_{sn}^{x}=1-\dfrac{1}{\sqrt{1+\dfrac{2\pi\eta^{2}}{\alpha}}}.
\end{align}

If the transverse noise is described by two orthogonal non-correlated components, 
the transition probability is averaged with a two-dimensional Gaussian
distribution:
\begin{align}\label{equ34}
P_{sn}^{xy}=\dfrac{1}{2\pi\eta^{2}}\int_{-\infty}^{\infty}dx\int_{-\infty}^{\infty}dyP_{\textmd{LZ}}(x,y)
\exp\Big(-\dfrac{x^{2}+y^{2}}{2\eta^{2}}\Big),
\end{align}
which after calculations acquires the form
\begin{align}\label{equ35}
P_{sn}^{xy}=1-\dfrac{1}{1+\dfrac{2\pi\eta^{2}}{\alpha}}.
\end{align}

The difference between Eqs.(\ref{equ33}) and (\ref{equ35}) is a consequence of the effective two-dimensional character of noise fluctuation spectrum in the
latter case and its one-dimensionality in the former case (see also Ref.\onlinecite{Kiselev}).

As expected, renormalization of the inter-level distance by a stochastic function considerably affects the generic picture of LZ
 transitions with off-diagonal coupling. Fast noise pours a large energy into the system during the crossing, destroys the memory of the system in the domain of strong couplings
identically distributing the system between the ground and excited states. We note, that in contrast to effects of the fast noise which does not change analytical properties of the LZ probability at extreme adiabatic limit $\alpha \to 0$, the two-component slow noise transfers the property of the LZ probability
from the Gaussian to the Lorentzian, thus making it an analytic function of $\alpha$ in this limit.

 The expansion method we exposed may allow one to formulate transition probabilities for finite times. For the case of $X$-noise, for example, the solution 
Eq.(\ref{equ19}) yields:
\begin{align}\label{equ36}
P_{sn}^{x}(t)=1-\dfrac{1}{\sqrt{1+\dfrac{2\pi\eta^{2}}{\alpha}\Big[F(t)+\ln W(t)\Big]}}.
\end{align}
In the limit $t\rightarrow\infty$, where $F(\infty)=1$ and $\ln W(\infty) \to 0$, we return to Eq.(\ref{equ33}).
One can do the same with the two-components transverse noise ($XY$-noise model) and get
\begin{align}\label{equ37}
P_{sn}^{xy}(t)=1-\dfrac{1}{1+\dfrac{2\pi\eta^{2}}{\alpha}\Big[F(t)+\ln W(t)\Big]}.
\end{align}
Similarly, the limit $t\rightarrow\infty$ brings us back to Eq.(\ref{equ35}). These solutions may now be interpreted
 in an interference pattern as they involve Fresnel-like integrals via $F(t)$.
\begin{figure}[t]
  \begin{center}
    \leavevmode
    \subfloat[]{%
      \label{fig5a}
      \includegraphics[width=4.3cm, height=35mm]{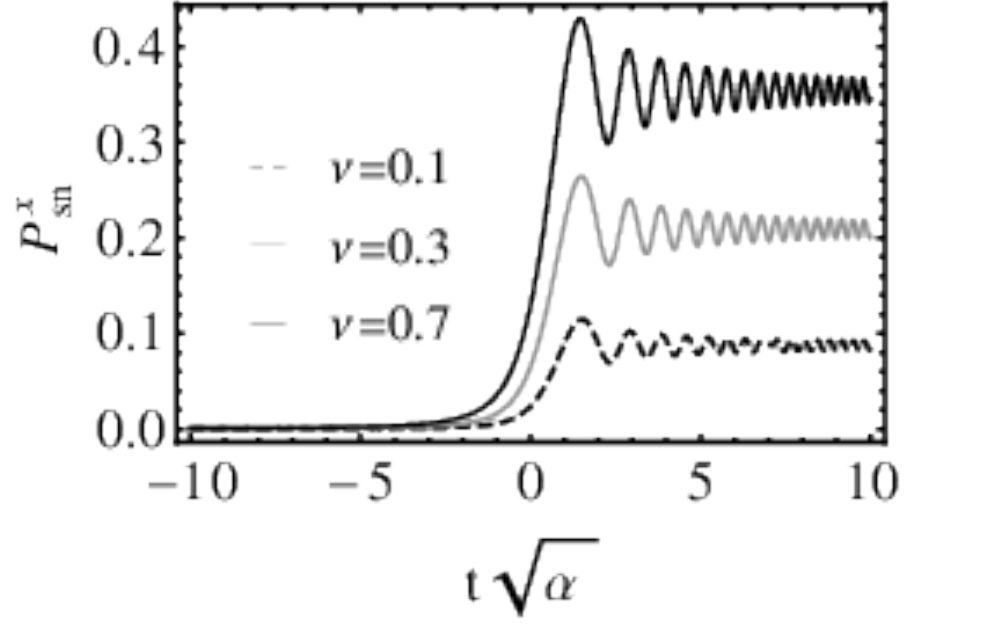}}
    \hspace{-6mm}
    \subfloat[]{%
      \label{fig5b}
      \includegraphics[width=4.3cm, height=35mm]{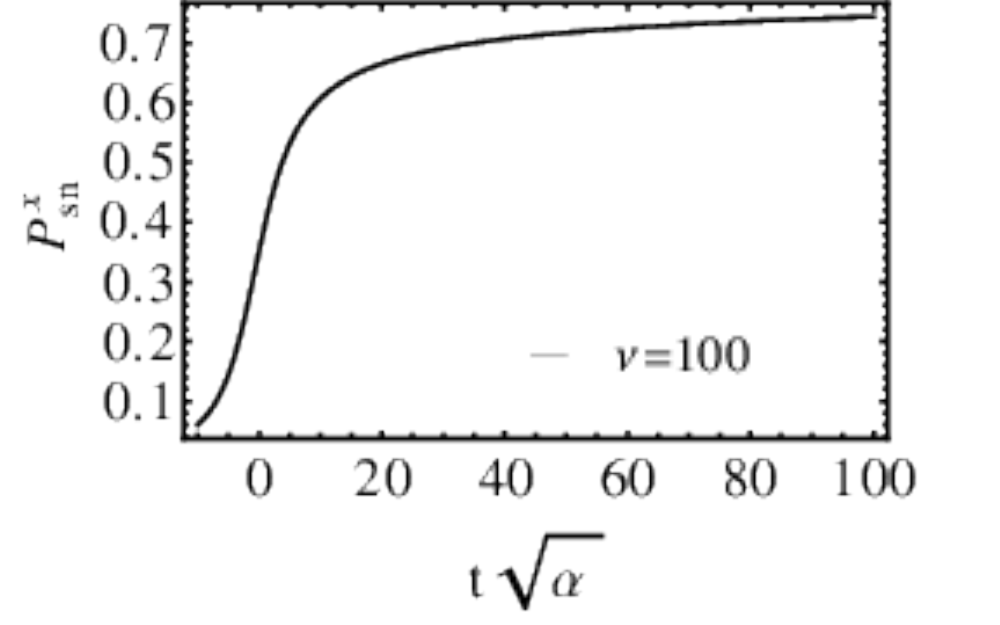}}
    \hspace{-6mm}
\subfloat[]{%
      \label{fig5c}
      \includegraphics[width=4.3cm, height=35mm]{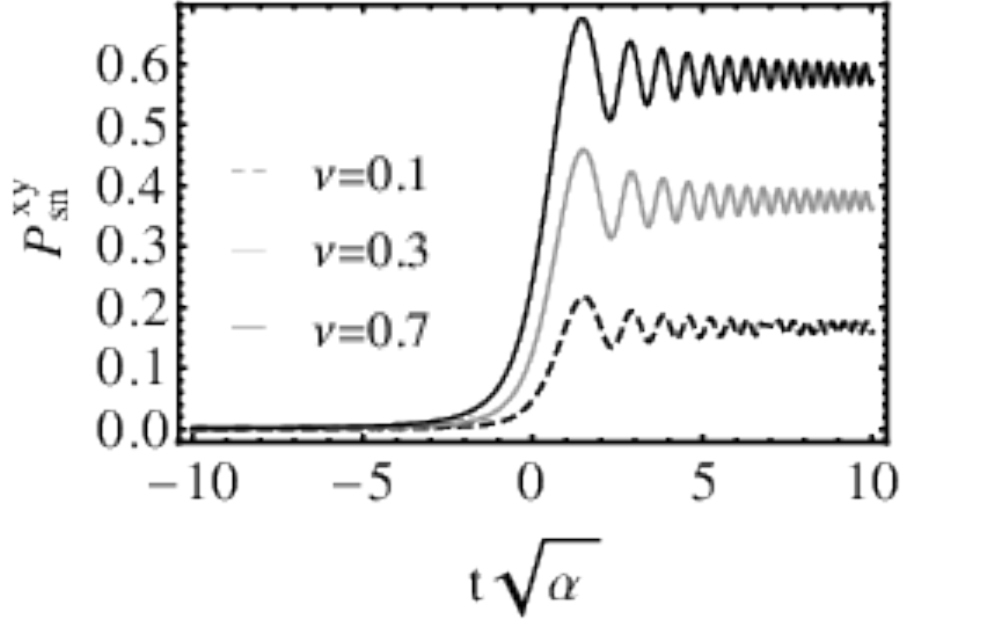}}
    \hspace{-6mm}
\subfloat[]{%
      \label{fig5d}
      \includegraphics[width=4.3cm, height=35mm]{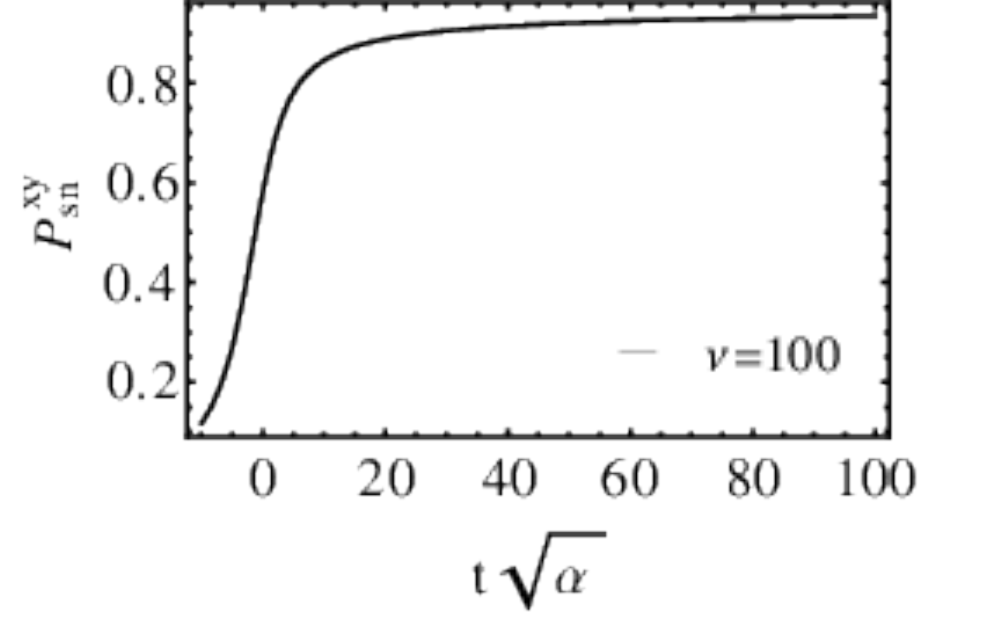}}
 \end{center}
\caption{Time evolution of the LZ transition probability in the diabatic basis of the two-level system in the presence of slow 
one- [(a) and (b)]  and two- [(c) and (d)]  transverse noises (see discussion in the text). 
 (a) and (c) represent the results of numerical calculations for the small- amplitude noise. The data for the large-amplitude noise are 
shown in  (b) and (d).
} \label{FIG5}
\end{figure}

$P_{sn}^{x}$ and $P_{sn}^{xy}$
have the same shape but $P_{sn}^{xy}$ is greater than $P_{sn}^{x}$ (see Fig. \ref{FIG5}).

\subsubsection*{Spin-$1/2$, in a constant off-diagonal field and a slow transverse random field}\label{Sec3.2.1}

\begin{figure}[b]
\begin{center}
 \includegraphics[width=7cm, height=4.5cm]{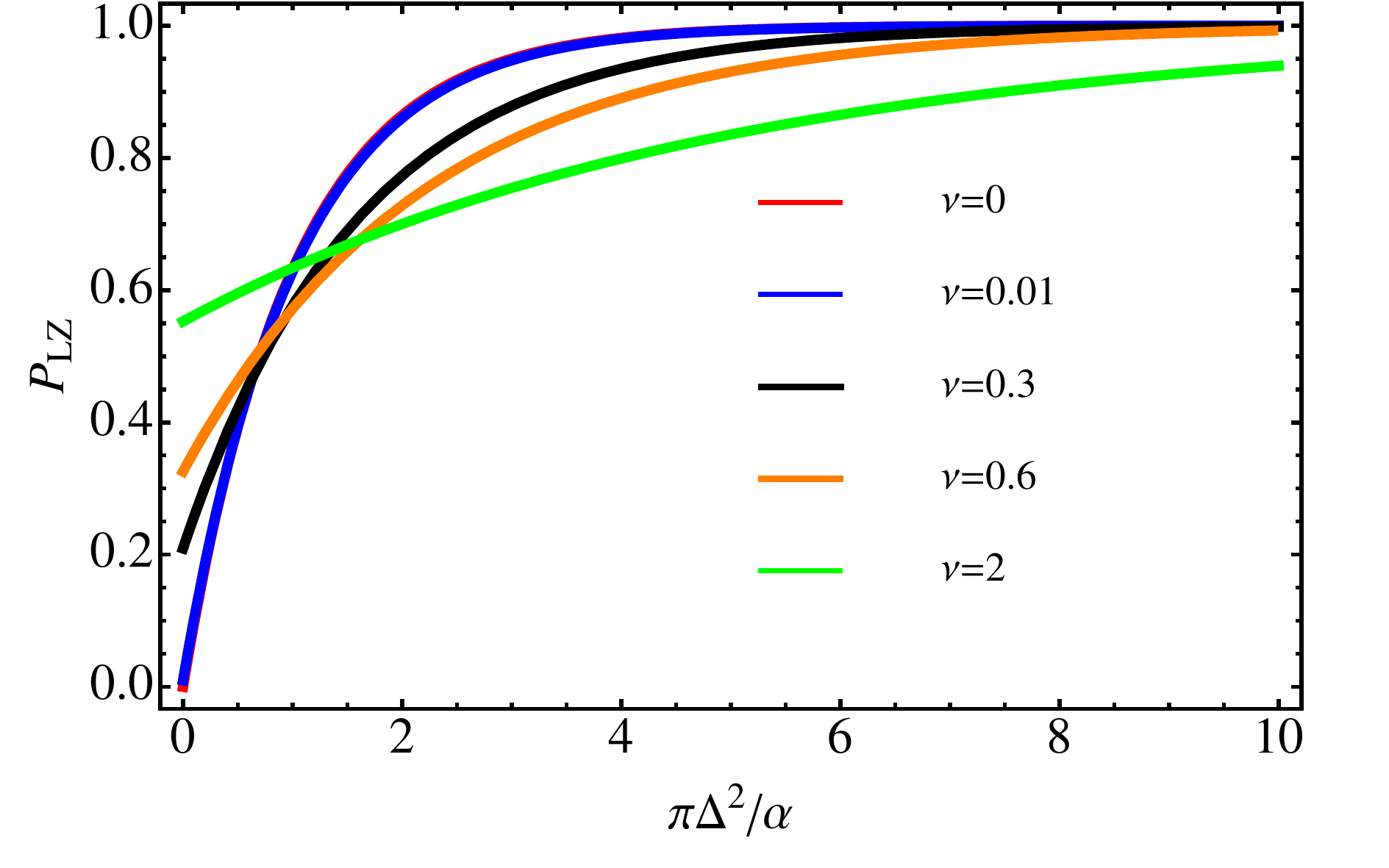}
\caption{(Color online) Landau-Zener transition probabilities for the two-level system at infinite time as a function of dimensionless parameter $\pi \Delta^2/\alpha$ in the presence of a one-component slow transverse noise. The parameter $\nu=\pi\eta^{2}/\alpha$ characterizes the noise amplitude.} \label{FIG6}
\end{center}
\end{figure}

We dealt up to this point with LZ transitions induced by the slow noise. This approach can be generalized straightforwardly
to the case where LZ transitions are induced by an external magnetic field (in the case of spin systems) or by an effective
 "field" associated with the finite transparency of the inter-well barrier in a double-well potential in cold gases. To describe this effect, one should
take into account the value of the stochastic field $f_{i}(t)$ as defined in Eq.(\ref{equa41}).
Then, averaging over slow one-component noise fluctuation, results in
\begin{align}\label{equ39}
\mathcal{P}_{sn}^{x}(t)=1-
\dfrac{\exp(-2\pi\lambda\Phi_{1}(t))}{\sqrt{1+\dfrac{2\pi\eta^{2}}{\alpha}\Big[F(t)+\ln W(t)\Big]}},
\end{align}
where the phase $\Phi_{\upsilon}(t)$ with $\upsilon=1,2$, expressed as
\begin{align}\label{equ40}
\Phi_{\upsilon}(t)=\dfrac{\Big[F(t)+\ln W(t)\Big]}{1+\dfrac{2\pi \upsilon\eta^{2}}{\alpha}\Big[F(t)+\ln W(t)\Big]}.
\end{align}
is due the local deviation of noise created in the $X$-direction. If the noise was also shifted along the $Y$-direction, we would have
 an additional phase such that the argument of the exponential in Eq.(\ref{equ39}) would be $\Phi_{\upsilon}(t)=\Phi^{x}_{\upsilon}(t)+\Phi^{y}_{\upsilon}(t)$. This would offer an access
to sum noises. Nonetheless, the choice we adopted has a great advantage in controlling noise fluctuations in a two-level system. For the meantime, the infinite time limit of Eq.(\ref{equ39}) suggests that:
\begin{align}\label{equ41}
\mathcal{P}_{sn}^{x}(\infty)=1-\dfrac{1}{\sqrt{1+\dfrac{2\pi \eta^{2}}{\alpha}}}\exp\Big\{-\dfrac{2\pi\lambda}{1+2\pi \eta^{2}/\alpha}\Big\}.
\end{align}
In the sudden limit,
\begin{align}\label{equ41a}
\mathcal{P}_{sn}^{x}(\infty)\approx P_{\textmd{LZ}}(\infty).
\end{align}
In the adiabatic limit,
\begin{align}\label{equ42}
\mathcal{P}_{sn}^{x}(\infty)=1-{\sqrt{\dfrac{\alpha}{2\pi \eta^{2}}}}\exp\Big\{-\dfrac{\Delta^{2}}{2\eta^{2}}\Big\}.
\end{align}
Thus, the argument of the exponent does not depend on the velocity.
Similarly, slow $XY$-noise in the presence of a constant magnetic field results in
\begin{align}\label{equ43}
\mathcal{P}_{sn}^{xy}(t)=1-
\dfrac{\exp(-2\pi\lambda\Phi_{1}(t))}{1+\dfrac{2\pi\eta^{2}}{\alpha}\Big[F(t)+\ln W(t)\Big]}.
\end{align}
Therefore,
\begin{align}\label{equ44}
\mathcal{P}_{sn}^{xy}(\infty)=1-\dfrac{1}{1+2\pi \eta^{2}/\alpha}\exp\Big\{-\dfrac{2\pi\lambda}{1+2\pi \eta^{2}/\alpha}\Big\},
\end{align}
and in the sudden limit,
\begin{align}\label{equ44a}
\mathcal{P}_{sn}^{xy}(\infty)\approx P_{\textmd{LZ}}(\infty),
\end{align}
while the adiabatic limit reads:
\begin{align}\label{equ45}
\mathcal{P}_{sn}^{xy}(\infty)=1-\dfrac{\alpha}{2\pi \eta^{2}}\exp\Big\{-\dfrac{\Delta^{2}}{2\eta^{2}}\Big\}.
\end{align}
A two-level system subjected to a small-amplitude ($\eta^{2}/\alpha\ll 1$) slow noise in the presence of a magnetic field is insensitive to noise structure for long-time asymptotic of transition probability.
 For such a setup, the magnetic field effects prevail on the noise and the pre-exponential factor is close to one.
These effects are supported by Fig.\ref{FIG6} where we plotted Eq.(\ref{equ41}) for $t=\infty$. We essentially show on Fig.\ref{FIG6} that the
 adiabatic addition of noise, considerably suppressed the previous tendency. 

Besides, by putting $\eta\rightarrow0$ in Eq.(\ref{equ41}) one immediately comes back to Eq.(\ref{equ19}), the LZ formula for finite times.

The averaging procedure described in details allows to calculate the LZ probability if, for example, the noise is fast in one of transverse directions and slow in another orthogonal direction. In that case,
one should first average the Bloch's equation over a fast realization and after solve the ``effective'' Bloch's equation in a given realization of slow fields. As a result, the fast noise contributes only to
the argument of LZ exponent, while the slow noise appears both in exponential and pre-exponential factors. Therefore, a numerical fit of experimental data\cite{Foletti2009, Foletti2010, Foletti2011}
could provide an information for both kind of noises without requiring additional measurements. 

By applying the {\em fluctuation dissipation theorem}, one can associate some effective temperature with an
equal time two-point correlation function as follows:
\begin{align}\label{equ46}
\langle f_{i}(t)^{2}\rangle=A\cdot T,
\end{align}
where $A$ and $T$ are respectively the coupling constant with the environment (model dependent) and the absolute temperature in the units $k_{B}=1$. In this frame, the solution
(\ref{equ45}) acquires the Arrhenius\cite{Levine} form
\begin{align}\label{equ47}
P_{sn}^{xy}(\infty)=1-\dfrac{\alpha}{2\pi A\cdot T}\exp\Big\{-\dfrac{E}{T}\Big\},
\end{align}
where $E=\Delta^{2}/2A $ is the activation energy\cite{kay}. The pre-exponential factor provides proper normalization of the distribution.

The theory of noise-induced LZ effect may be extended to the multilevel LZ problems, where more complicated patterns of transient
 oscillations in the tunneling probability of transition from the initial to the final state of a nanosystem with nontrivial dynamical symmetry
is expected.

\section{Basic relations for three-level systems}\label{Sec4}
\subsection{Schr\"odinger  spin-$1$ picture}\label{Sec4.1}
Consider the LZ transition in a three-level spin-1 system with the upper and lower levels described respectively by
$C_{1}^{(1)}(t)$ and $C_{2}^{(1)}(t)$ that characterize, respectively, the eigenenergy states $E_{\pm}(t)=\pm2(\alpha^{2}t^{2}+\Delta^{2})^{1/2}$.
 The wave function $C_{0}^{(1)}(t)$ characterizing the middle level corresponds to the eigenenergy state $E(t)=0$. This eigenenergy state does not
evolve in time and the transitions between neighboring energy levels are allowed.

The operator $S^{z}$  is diagonal in its eigen representation
and has the eigen-values $-1$, $0$ and $+1$ as diagonal elements that match respectively the states $\lvert1\rangle$, $\lvert0\rangle$ and $\lvert2\rangle$
 which form avoided-level crossing points.
From Eqs.(\ref{equ1}) and (\ref{equ3}), we arrive at a system of three decoupled differentials equations for the states
$C_{1}^{(1)}(t)$, $C_{0}^{(1)}(t)$ and $C_{2}^{(1)}(t)$. The first pair of equations for the states with minimal/maximal projection to $z$-axis is:
\begin{align}\label{equ48a}
\dfrac{d^{3}}{dz^{3}}C_{1}^{(1)}(z)+(4i\lambda-2-z^{2})\dfrac{d}{dz}C_{1}^{(1)}(z)-zC_{1}^{(1)}(z)=0,
\end{align}
\begin{align}\label{equ48c}
\dfrac{d^{3}}{dz^{3}}C_{2}^{(1)}(z)+(4i\lambda+2-z^{2})\dfrac{d}{dz}C_{2}^{(1)}(z)-zC_{2}^{(1)}(z)=0.
\end{align}
The dynamics of the middle level is independently derived and is governed by a third order linear differential equation
of the form (\ref{equ25}). For sake of consistency it is presented here as follows:
\begin{align}\nonumber\label{equ48b}
\dfrac{d^{3}}{dz^{3}}C_{0}^{(1)}(z)-\dfrac{1}{z}\dfrac{d^{2}}{dz^{2}}C_{0}^{(1)}(z)- 4\Big[\dfrac{i\lambda}{z}C_{0}^{(1)}(z)\\&&\hspace{-3cm}+(\dfrac{z^{2}}{4}-i\lambda)\dfrac{d}{dz}C_{0}^{(1)}(z)\Big]=0.
\end{align}
In obtaining these equations, no assumptions on the initial preparation of the system  have been adopted.
One may arbitrary select a particular ground state and ask questions about the probability to find the system on the excited states.

Considering Eqs.(\ref{equ16}) and (\ref{equ25}) then Eq.(\ref{equ48b}) may take the form:
\begin{align}\label{equ48b1}
\dfrac{d }{dz}C_{0}^{(1)}(z)=-4i\lambda\int_{-\infty}^{z}dz_{1}\cosh\Big[\dfrac{1}{2}(z^{2}-z_{1}^{2})\Big]C_{0}^{(1)}(z_{1}).
\end{align}
To find the solutions of Eqs.(\ref{equ48a})- and (\ref{equ48c}), it would be instructive to know that the square of the Weber's functions in Eqs.(\ref{equ4}) and (\ref{equ5}) satisfies the third-order differential equation\cite{Mitra}:
\begin{align}\nonumber\label{equ48d}
\dfrac{d^{3}}{dz^{3}}\Big[C_{1}^{(1/2)}(z)\Big]^{2}+(4i\lambda-2-z^{2})\dfrac{d}{dz}\Big[C_{1}^{(1/2)}(z)\Big]^{2}\\-z\Big[C_{1}^{(1/2)}(z)\Big]^{2}=0,
\end{align}
\begin{align}\nonumber\label{equ48e}
\dfrac{d^{3}}{dz^{3}}\Big[C_{2}^{(1/2)}(z)\Big]^{2}+(4i\lambda+2-z^{2})\dfrac{d}{dz}\Big[C_{2}^{(1/2)}(z)\Big]^{2}\\-z\Big[C_{2}^{(1/2)}(z)\Big]^{2}=0.
\end{align}
For consistency, we will express all our solutions through the Weber's function. The solutions to these equations are $D_{i\lambda}^{2}(z)$, $D_{-i\lambda-1}^{2}(iz)$, $D_{-i\lambda-1}^{2}(-iz)$ or any product of any of the functions    $D_{i\lambda}(z)$, $D_{-i\lambda-1}(iz)$, and $D_{-i\lambda-1}(-iz)$\cite{Mitra}.
With given initial conditions, these give the basis to the solutions of Eqs.(\ref{equ48a}) and (\ref{equ48c}).

Consider the case when the initial conditions are:
\begin{align}\label{equ48fai}
C_{1}^{(1)}(-\infty)=1, \quad C_{0}^{(1)}(-\infty)=0 \quad \textmd{and} \quad C_{2}^{(1)}(-\infty)=0,
\end{align}
then $C_{1}^{(1)}(t)=B_{-}D_{-i\lambda}^{2}(-iz)$ and $C_{2}^{(1)}(t)=B_{+}D_{-i\lambda-1}^{2}(-iz)$.
Here, $B_{\pm}$ are normalization factors. We establish relations between the wave functions of the triplet $C_{1}^{(1)}(t)$, $C_{0}^{(1)}(t)$ , $C_{2}^{(1)}(t)$
and the doublet $C_{1}^{(1/2)}(t)$, $C_{2}^{(1/2)}(t)$ states considering in addition the normalization condition $\sum_{m=-S}^{S}\lvert C_{m}^{(S)}(t)\rvert^{2}=1$:
\begin{align}\label{equ48f}
C_{1}^{(1)}(t)=\Big[C_{1}^{(1/2)}(t)\Big]^{2},
\end{align}
\begin{align}\label{equ48g}
C_{0}^{(1)}(t)=\sqrt{2}C_{1}^{(1/2)}(t)C_{2}^{(1/2)}(t),
\end{align}
\begin{align}\label{equ48h}
C_{2}^{(1)}(t)=\Big[C_{2}^{(1/2)}(t)\Big]^{2}.
\end{align}
Considering the conditions
\begin{align}\label{equ48e1}
C_{1}^{(1)}(-\infty)=0,\quad C_{0}^{(1)}(-\infty)=1\quad \textmd{and} \quad C_{2}^{(1)}(-\infty)=0,
\end{align}
 Eq.(\ref{equ48b}) or its integral-differential form (\ref{equ48b1}) is isomorphic to Eq.(\ref{equ25}):
\begin{align}\nonumber\label{equ49}
 C_{0}^{(1)}(t)=-\lambda e^{-\pi \lambda/2}\Big[\lvert D_{-i\lambda-1}(-i\mu t)\rvert^{2}\\&&\hspace{-2cm} -\dfrac{1}{\lambda}\lvert D_{-i\lambda}(-i\mu t)\rvert^{2} \Big].
\end{align}
From conditions $C_{1}^{(1)}(-\infty)=C_{2}^{(1)}(-\infty)=0$,  the solutions of Eqs. (\ref{equ48a}) and (\ref{equ48c}) satisfy the following equations:
\begin{align}\label{equ50a}
C_{1}^{(1)}(t)=\sqrt{2\lambda}e^{-i\pi/4}\int^{z}_{-\infty }\exp\Big(\frac{1}{2}[z^{2}-z^2_{1}]\Big)C_{0}^{(1)}(z_{1})dz_{1},
\end{align}
\begin{align}\label{equ50b}
C_{2}^{(1)}(t)=\sqrt{2\lambda}e^{-i\pi/4}\int^{z}_{-\infty }\exp\Big(-\frac{1}{2}[z^{2}-z^2_{1}]\Big)C_{0}^{(1)}(z_{1})dz_{1},
\end{align}
and
\begin{align}\label{equ50c}
\dfrac{d}{dz}C_{0}^{(1)}(z)=\sqrt{2\lambda}\Big(C_{1}^{(1)}(z)+C_{2}^{(1)}(z)\Big)e^{-i\pi/4}.
\end{align}
Substituting Eq.(\ref{equ49}) into Eqs.(\ref{equ50a}) and (\ref{equ50b}) and considering Eq.(2.27) then this yields
\begin{align}\nonumber\label{equ51}
C_{1}^{(1)}(t)=\sqrt{2\lambda}\exp\Big(i\varphi'-\dfrac{ i\pi }{4}-\dfrac{\pi\lambda}{2}\Big)\\
&&\hspace{-3.5cm} D_{-i\lambda }(-i\mu t)[D_{-i\lambda-1}(-i\mu t)]^{*},
\end{align}\\
\begin{align}\nonumber\label{equa51}
C_{2}^{(1)}(t)=-\sqrt{2\lambda}\exp\Big(i\varphi' +\dfrac{ i\pi }{4}-\dfrac{\pi\lambda}{2}\Big)\\
&&\hspace{-3.5cm}[D_{-i\lambda}(-i\mu t)]^{*}D_{-i\lambda- 1}(-i\mu t).
\end{align}
Here, $\varphi'$ is an arbitrary phase factor. The above permits to achieve LZ transition probabilities expressed through the following :
\begin{align}\label{equa52}
 C_{1}^{(1)}(t)=-\sqrt{2}C_{1}^{(1/2)}(t)C_{2}^{(1/2)*}(t),
\end{align}
\begin{align}\label{equa53}
 C_{0}^{(1)}(t)=\lvert C_{1}^{(1/2)}(t)\rvert^{2}-\lvert C_{2}^{(1/2)}(t)\rvert^{2},
\end{align}
 \begin{align}\label{equa54}
 C_{2}^{(1)}(t)=\sqrt{2}C_{1}^{(1/2)*}(t)C_{2}^{(1/2)}(t).
\end{align}
For completeness, the solution of (\ref{equ48a})-(\ref{equ48b}) with initial conditions
 \begin{align}\label{equa54a}
C_{1}^{(1)}(-\infty)=0, \quad C_{0}^{(1)}(-\infty)=0 \quad \textmd{and} \quad C_{2}^{(1)}(-\infty)=1
\end{align}
can be found with the help of Eqs. (\ref{equa52})-(\ref{equa54}).
It is instructive to know that the three-level system for $S=1$ possesses an additional symmetry level that imitates a particle-hole $SU(2)$ symmetry group \cite{Kiselev}.
A transition matrix  for $S=1$ is then constructed as follows:
\begin{widetext}
\begin{align}\label{equa55a}
\hat{\mathcal{U}}^{\textmd{LZ}}(t)=
\begin{bmatrix}
 \Big[C_{1}^{(1/2)}(t)\Big]^{2} &&  \sqrt{2}C_{1}^{(1/2)}(t)C_{2}^{(1/2)}(t)&&\Big[C_{2}^{(1/2)}(t)\Big]^{2}\\\\
 -\sqrt{2}C_{1}^{(1/2)}(t)C_{2}^{(1/2)*}(t) &&  \lvert C_{1}^{(1/2)}(t)\rvert^{2}-\lvert C_{2}^{(1/2)}(t)\rvert^{2} && \sqrt{2}C_{1}^{(1/2)*}(t)C_{2}^{(1/2)}(t)\\\\
 \Big[C_{2}^{(1/2)*}(t)\Big]^{2} && -\sqrt{2}C_{1}^{(1/2)*}(t)C_{2}^{(1/2)*}(t)&& \Big[C_{1}^{(1/2)*}(t)\Big]^{2}
\end{bmatrix}
\end{align}
\end{widetext}
The transition  matrix in Eq.(\ref{equa55a}) is thus a generalization of the result in Ref. \onlinecite{pok}.
Here, the matrix element $\mathcal{U}^{\textmd{LZ}}_{nm}(t)$ is the transition amplitude for the transition from the diabatic state $\rvert n \rangle$ to
$\rvert m \rangle$. Applying the condition $t=\infty$ to our generalized results yields
exactly those in Ref. \onlinecite{pok} for all transition matrix elements (see Table.\ref{TAB1}).

\begin{figure}[]
  \begin{center}
    \leavevmode
    \subfloat[]{%
      \label{fig7a}
      \includegraphics[width=4.3cm, height=35mm]{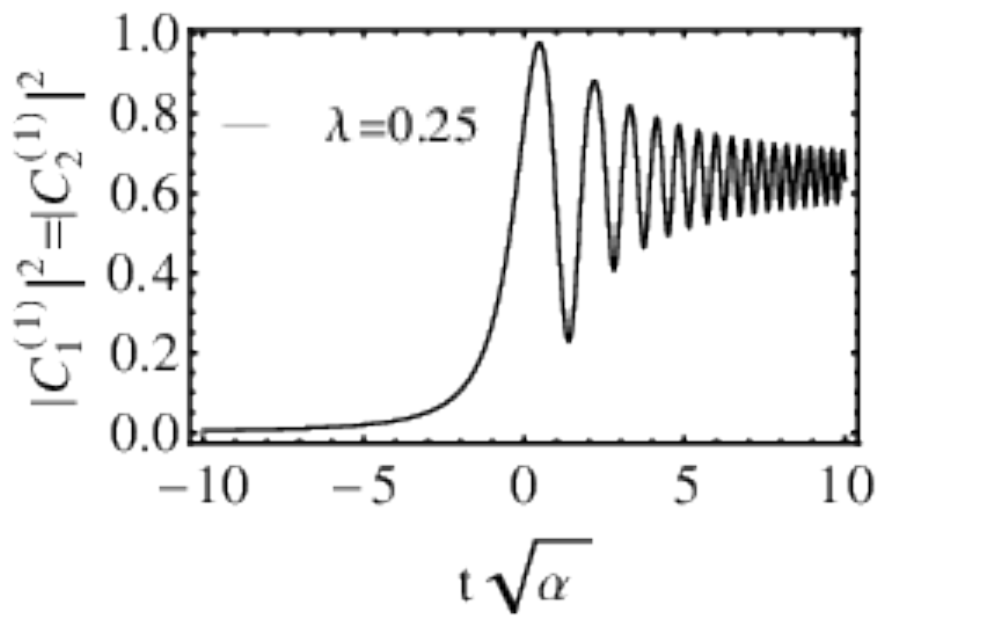}}
    \hspace{-6mm}
    \subfloat[]{%
      \label{fig7b}
      \includegraphics[width=4.3cm, height=35mm]{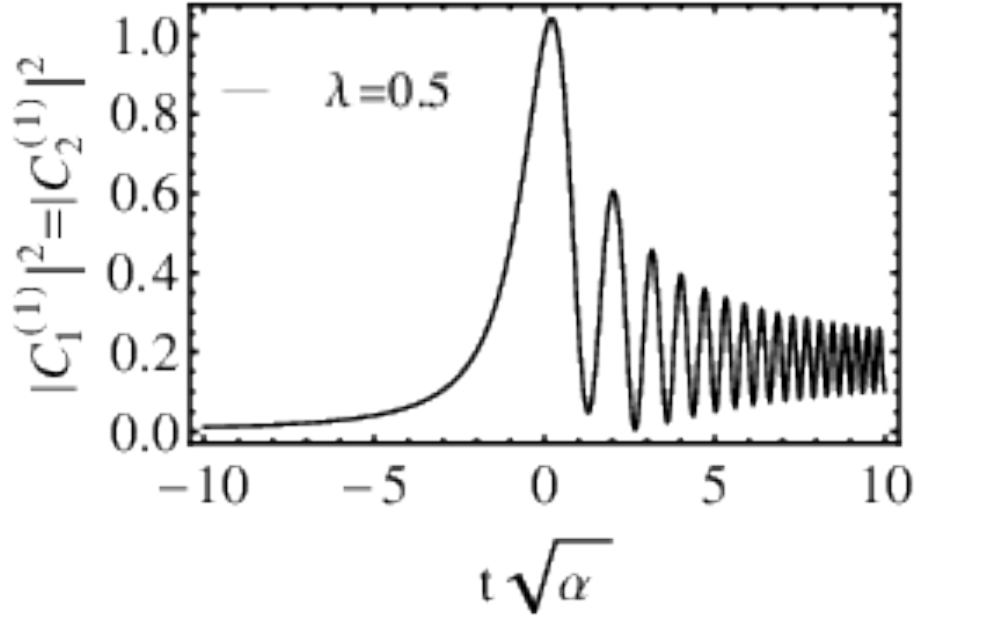}}
    \hspace{-6mm}
\subfloat[]{%
      \label{fig7c}
      \includegraphics[width=4.3cm, height=35mm]{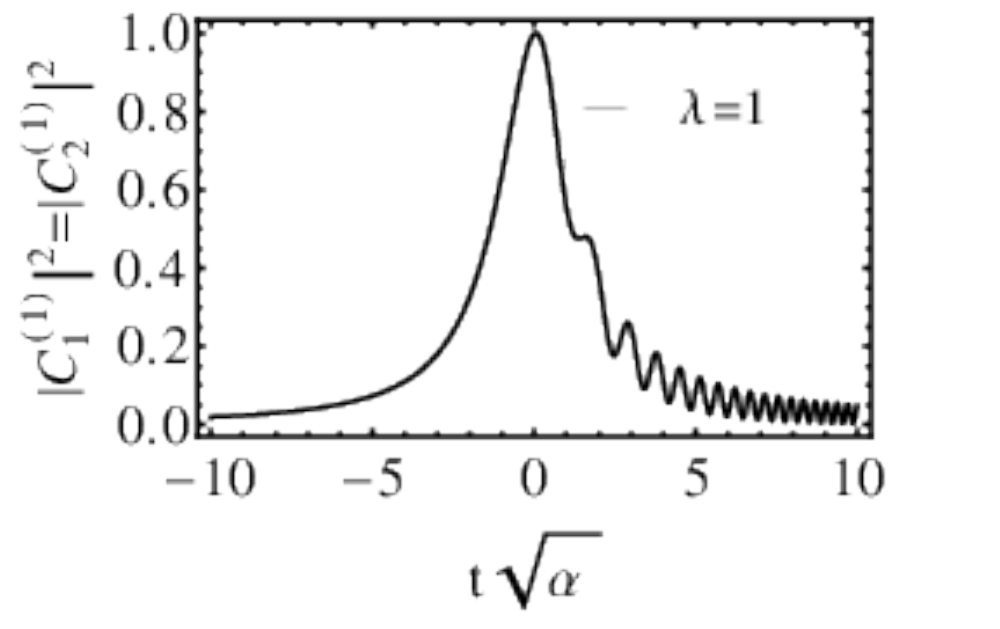}}
    \hspace{-6mm}
\subfloat[]{%
      \label{fig7d}
      \includegraphics[width=4.3cm, height=35mm]{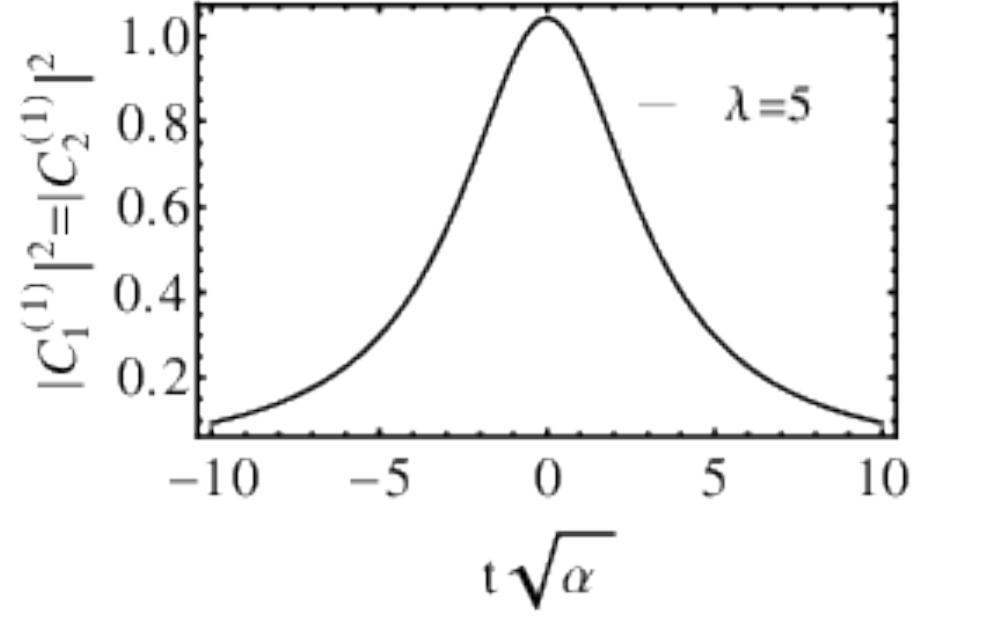}}
 \end{center}
\caption{Time evolution of the LZ transition probability in the diabatic basis of the three-level system.
 (a) and (d) show the numerical results for rapid and slow LZ drive respectively. The results on intermediate regime are presented in (b) and (c). The LZ parameter $\lambda=\Delta^2/2\alpha$. The time is in the units of $1/\sqrt{\alpha}$.} \label{FIG6a}
\end{figure}

We find numerically the dynamical evolution of the model (\ref{equ2}) for $S=1$ by solving the Schr\"odinger equation for the amplitudes
$C_{n}^{(1)}(t)$, $(n=1,0,2)$
then we plot the population $\rvert C_{n}^{(1)}(t)\lvert^{2}$ considering conditions (\ref{equ48e}). These results are depicted on Fig.\ref{FIG6a}.
As foreseen, the populations $\rvert C_{1}^{(1)}(t)\lvert^{2}$ and $\rvert C_{2}^{(1)}(t)\lvert^{2}$
 are identically distributed on levels defined by the states $\rvert 1 \rangle$ and $\rvert 2 \rangle$ .

Figures \ref{FIG6a}(a)-\ref{FIG6a}(d) show how population of each of the levels changes with the LZ parameter.
Taking limits in Eq.(\ref{equa55a}) as $t\rightarrow\infty$ achieves the results in Ref. \onlinecite{Caroll1986} (see Table.\ref{TAB1}).
\begin{table}{}
\begin{tabular}{lcr}
\hline
\hline
Initial occupation for $t=-\infty$ & $\rvert$ Final occupation for $t=\infty$ \\
\hline\\
1 & $ e^{-4\pi\lambda}$ \\ \\ 0 & $2[e^{-2\pi\lambda}-e^{-4\pi\lambda}]$\\ \\ 0 & $[1-e^{-2\pi\lambda}]^{2}$\\
\\ \\
0 & $2[e^{-2\pi\lambda}-e^{-4\pi\lambda}]$ \\ \\ 1 & $[1-2e^{-2\pi\lambda}]^{2}$\\ \\ 0 & $2[e^{-2\pi\lambda}-e^{-4\pi\lambda}]$\\
\\ \\
0 & $[1-e^{-2\pi\lambda}]^{2}$ \\ \\ 0 & $2[e^{-2\pi\lambda}-e^{-4\pi\lambda}]$\\ \\ 1 & $e^{-4\pi\lambda}$\\
\hline
\hline
\end{tabular}
\caption{{\small Landau-Zener transition probabilities in the three-level system.}}\label{TAB1}
\end{table}

\subsection{Bloch spin-$1$ picture}\label{Sec4.2}

The focus in this heading is exclusively on $P_{01}(t)=|{\mathcal U}^{\textmd{LZ}}_{01}(t)|^2$ and $P_{00}(t)=|{\mathcal U}^{\textmd{LZ}}_{00}(t)|^2$.
Here ${\mathcal U}^{\textmd{LZ}}_{01}(t)$  and ${\mathcal U}^{\textmd{LZ}}_{00}(t)$  are matrix elements of the time evolution
operator $3\times3$ matrix defined in Eq.(\ref{equa55a}).

The dynamics of the system is governed by nine equations for the components of the $3\times3$ density matrix. This is reduced to a set of six equations due to the symmetry
of the levels. Knowledge of two of the matrix elements is enough to compute the other matrix elements considering the condition for the probability conservation.
In this regard, we reduce the problem to a set of two equations for
$\hat{\rho}_{+}(t)=\hat{\rho}_{11}(t)-\hat{\rho}_{00}(t)$ and $\hat{\rho}_{-}(t)=\hat{\rho}_{22}(t)-\hat{\rho}_{00}(t)$.
Here, the indices $1, 0$ and $2$ denote the three crossing levels.
It would be convenient to express the probabilities $P_{01}(t)$ and $P_{00}(t)$ in exponential form as we did in the preceding section for LZ spin-$1/2$ probabilities.
Equation (\ref{equ49})  establishes a relation between the occupation probability $P_{00}(t)$ and the LZ transition probability $P_{\textmd{LZ}}(t)$:
\begin{align}\label{equ56fai}
P_{00}(t)=(2P_{\textmd{LZ}}(t)-1)^{2}.
\end{align}
The normalization of probabilities helps one to express the finite tunneling time probabilities as:
\begin{align}\nonumber\label{equ56}
P_{01}(t)=2\Big[\exp\Big\{-2\pi\lambda[F(t)+\ln W(t)]\Big\}-\\
\exp\Big\{-4\pi\lambda[F(t)+\ln W(t)]\Big\}\Big],
\end{align}
and
\begin{align}\label{equ57}
P_{00}(t)=\Big[1-2\exp\Big\{-2\pi\lambda[F(t)+\ln W(t)]\Big\}\Big]^{2}.
\end{align}
This will aid to derive the LZ transition probabilities in the slow noise approximation.
This paper considers the transition probabilities in the fast and slow noise approximations.
Detailed calculations for relevant equations are found in Appendix \ref{App2}.

We show the correspondence between the Schr\"odinger and Bloch approaches. Schr\"odinger dynamics of $\mathcal{N}$-level systems describing spin $S=(\mathcal{N}-1)/2$ may be expressed through
a set of $\mathcal{N}$ coupled first-order linear differential equations (LDE). These equations represent $\mathcal{N}$ independent $\mathcal{N}$-th order LDE.

The time evolution operator expressed through Jacobi $\mathcal{N}-1$ order polynomials\cite{QM, Kenmoe} may be constructed on the basis of the $SU(2)$ group with fundamental  spinors.
Therefore, the solution of $\mathcal{N}$-th order LDE is expressed through the   $(\mathcal{N}-1)$-fold product of Weber's functions. The Bloch dynamics of the spin $S$ is based on one vector
and $2S-1$ tensor Bloch equations. This is due to the fact that the density matrix has $2S$ conservable values\cite{pok}.

\section{Transverse noise in the spin-$1$ Landau-Zener theory}\label{Sec5}

In this section we evaluate the tunneling probabilities for the case when the inter-level distance between the states of a three-level system is renormalized by a random classical field. Similar studies were performed by Pokrovsky\cite{pok2003} with restriction to fast noise. To the best of our knowledge the slow noise approximation has not yet been investigated for three-level systems. So, we study the LZ transition probabilities for the three-level system in the slow noise approximation by applying an ensemble averaging over all possible noise realizations.
The procedure to obtain the equation of motion for the density matrix describing transitions in three-level systems imitate that of $S=1/2$ in sec. \ref{Sec2} of 
this paper (details of the procedure can be found in Appendix \ref{App2}). From Eqs. (\ref{Equ1})-(\ref{Equ5}) one gets:

\begin{widetext}
\begin{align}\label{equ58}
\dfrac{d\hat{\rho}_{+}(t)}{dt}=-4\int_{-\infty}^{t}\Theta_{+}(t)\Theta_{-}(t_{1})
\cos\Big[\alpha(t^{2}-t_{1}^{2})\Big]\hat{\rho}_{+}(t_{1})dt_{1}-2\int_{-\infty}^{t}\Theta_{+}(t)\Theta_{-}(t_{1})
\cos\Big[\alpha(t^{2}-t_{1}^{2})\Big]\hat{\rho}_{-}(t_{1})dt_{1}+\Phi,
\end{align}
\begin{align}\label{equ59}
\dfrac{d\hat{\rho}_{-}(t)}{dt}=-2\int_{-\infty}^{t}\Theta_{+}(t)\Theta_{-}(t_{1})\cos\Big[\alpha(t^{2}-t_{1}^{2})\Big]
\hat{\rho}_{+}(t_{1})dt_{1}-4\int_{-\infty}^{t}\Theta_{+}(t)\Theta_{-}(t_{1})
\cos\Big[\alpha(t^{2}-t_{1}^{2})\Big]\hat{\rho}_{-}(t_{1})dt_{1}+\Psi.
\end{align}
\end{widetext}
Here, $\Theta_{\pm}(t)=(\Theta^{x}(t)\pm i\Theta^{y}(t))/\sqrt{2}$. Equations (\ref{equ58}) and (\ref{equ59}) are obtained considering the functions $\Theta_{+}(t)$ and $\Theta_{-}(t)$ to be noise fields.  $\Phi$  and $\Psi$ are functions of $\Theta_{+}(t)\Theta_{+}(t_{1})$ and $\Theta_{-}(t)\Theta_{-}(t_{1})$. It is instructive to note that the averages $\langle \Theta_{+}(t)\Theta_{+}(t_{1})\rangle$ and $\langle \Theta_{-}(t)\Theta_{-}(t_{1})\rangle$ vanish as Gaussian correlators and consequently $\Phi$  and $\Psi$ will not contribute to the transition probabilities. If $\Theta_{+}(t)$ and $\Theta_{-}(t)$ are not noise correlated then this is not applicable and the components $\hat{\rho}_{12}(t)$ and $\hat{\rho}_{21}(t)$ will enter the expression of the density matrix.

We verify if the method employed in Sec. \ref{Sec2} for the spin-$1/2$ LZ transition relates a third-order differential equation considering $\hat{\rho}_{12}^{(0)}(t)$ and $\hat{\rho}_{21}^{(0)}(t)$:
\begin{widetext}
\begin{align}\nonumber\label{equation60}
\dfrac{d}{dt}\hat{\rho}_{+}^{(0)}(t)=-8\Delta^{2}\int_{-\infty}^{t}\cos\Big[\alpha(t^{2}-t_{1}^{2})\Big]\hat{\rho}_{+}^{(0)}(t_{1})dt_{1}
-4\Delta^{2}\int_{-\infty}^{t}\cos\Big[\alpha(t^{2}-t_{1}^{2})\Big]\hat{\rho}_{-}^{(0)}(t_{1})dt_{1}\\&&\hspace{-12cm}-6\Delta^{2}\int_{-\infty}^{t}\exp\Big[i\alpha(t^{2}-t_{1}^{2})\Big]\hat{\rho}_{21}^{(0)}(t_{1})dt_{1}
-6\Delta^{2}\int_{-\infty}^{t}\exp\Big[-i\alpha(t^{2}-t_{1}^{2})\Big]\hat{\rho}_{12}^{(0)}(t_{1})dt_{1},
\end{align}
\begin{align}\nonumber\label{equation61}
\dfrac{d}{dt}\hat{\rho}_{-}^{(0)}(t)=-4\Delta^{2}\int_{-\infty}^{t}\cos\Big[\alpha(t^{2}-t_{1}^{2})\Big]
\hat{\rho}_{+}^{(0)}(t_{1})dt_{1}-8\Delta^{2}\int_{-\infty}^{t}\cos\Big[\alpha(t^{2}-t_{1}^{2})\Big]\hat{\rho}_{-}^{(0)}(t_{1})dt_{1}\\&&\hspace{-12cm}-6\Delta^{2}\int_{-\infty}^{t}\exp\Big[i\alpha(t^{2}-t_{1}^{2})\Big]\hat{\rho}_{21}^{(0)}(t_{1})dt_{1}
-6\Delta^{2}\int_{-\infty}^{t}\exp\Big[-i\alpha(t^{2}-t_{1}^{2})\Big]\hat{\rho}_{12}^{(0)}(t_{1})dt_{1}.
\end{align}
\end{widetext}
It is less obvious to derive a third order differential equation for the matrices:
\begin{align}\label{equation63}
&&\hspace{-.6cm}
 \hat{\rho}^{(0)}(t)=
\begin{bmatrix}
\hat{\rho}_{+}^{(0)}(t)\\
\hat{\rho}_{-}^{(0)}(t)
\end{bmatrix}
\quad
\textmd{and}
\quad
\hat{\mathcal{Q}}^{(0)}(t)=
 \begin{bmatrix}
\hat{\rho}_{21}^{(0)}(t)\\
\hat{\rho}_{12}^{(0)}(t)
\end{bmatrix},
\end{align}
as we did for the spin $S=1/2$. The matrix elements in the second matrix of equation (\ref{equation63}) generated by the last two terms in Eqs.(\ref{equation60}) and (\ref{equation61}) might be viewed as external sources for a homogeneous matrix element equation of the form (\ref{equ16}). However, an appropriate choice of variables leads to the non homogeneous equation:
\begin{align}\nonumber\label{equation62}
&&\hspace{-1.5cm}\dfrac{d}{dt}\hat{\rho}^{(0)}(t)=-4\Delta^{2}_{M}\int_{-\infty}^{t}\cos\Big[\alpha(t^{2}-t_{1}^{2})\Big]\hat{\rho}^{(0)}(t_{1})dt_{1}
\\&&\hspace{-4.5cm}-6\Delta^{2}\hat{\mathcal{J}}(t)\hat{\mathcal{Q}}^{(0)}(t).
\end{align}
Obtaining this, we define corresponding functions in the absence of noise as follows as in Eq.(\ref{equation63}). In Eq.(\ref{equation62}), 
the square of the inter-level distance is given by the following matrix
\begin{align}\label{Equation65}
\Delta^{2}_{M}=2\Delta^{2}
\begin{pmatrix}
1 & 1/2 \\ 1/2 & 1
\end{pmatrix}.
\end{align}
The operator $\hat{\mathcal{J}}(t)$ acts onto the subspace $\mathcal{S}_{b}$ of the vector $\hat{\mathcal{Q}}^{(0)}(t)$  and generates the last two terms in Eqs.(\ref{equation60}) and (\ref{equation61}). 
The Lie algebra associated with the time derivative properties of $\hat{\mathcal{J}}(t)$ can be defined from the relation
\begin{align}\nonumber\label{Equation63}
\hat{\mathcal{J}}(t)\hat{\rho}_{12}^{(0)}(t)=\int_{-\infty}^{t}\cos\Big[\alpha(t^{2}-t_{1}^{2})\Big]\textmd{Re}[\hat{\rho}_{12}^{(0)}(t_{1})]dt_{1}
\\-i\int_{-\infty}^{t}\sin\Big[\alpha(t^{2}-t_{1}^{2})\Big]\textmd{Im}[\hat{\rho}_{12}^{(0)}(t_{1})]dt_{1}.
\end{align}
Eq.(\ref{Equation63}) shows $\hat{\mathcal{J}}(t)$ in action into $\mathcal{S}_{b}$. The action of $\hat{\mathcal{J}}(t)$ is symmetric
 so that $\hat{\mathcal{J}}(t)\hat{\rho}_{21}^{(0)}(t)=\hat{\mathcal{J}}(t)\hat{\rho}_{12}^{(0)}(t)$.
Obviously, the operator $d/dt$ and its higher orders $d^{q}/dt^{q}$ ($q>1$) have non-zero actions in $\mathcal{S}_{b}$ .
The simultaneous actions $\frac{d}{dt}\circ\hat{\mathcal{J}}(t)$ and $\frac{d^{2}}{dt^{2}}\circ\hat{\mathcal{J}}(t)$ including $\frac{d^{q}}{dt^{q}}\circ\hat{\mathcal{J}}(t)$ are then also
defined in $\mathcal{S}_{b}$.

By evaluating $\frac{d}{dt}\hat{\mathcal{J}}(t)\hat{\rho}_{12}^{(0)}(t)$ and $\frac{d^{2}}{dt^{2}}\hat{\mathcal{J}}(t)\hat{\rho}_{12}^{(0)}(t)$ and considering
the initial condition $\hat{\rho}_{12}^{(0)}(-\infty)=0$ we define equivalent initial conditions helpful for further purposes. It can be verified that,
\begin{align}\label{equation64}
\dfrac{d}{dt}\hat{\mathcal{J}}(t)\hat{\rho}_{12}^{(0)}(t)\mid_{t=-\infty}=\hat{\mathcal{J}}(t)\hat{\rho}_{12}^{(0)}(t)\mid_{t=-\infty}=0.
\end{align}
These properties are directly applicable to $\hat{\rho}_{21}^{(0)}(t)$ as $\hat{\rho}_{21}^{(0)}(-\infty)=0$.
 We omitted the symbol $\circ$ keeping in mind the ordering of actions in $\mathcal{S}_{b}$, namely, $\hat{\mathcal{J}}(t)$ first passes and $d^{q}/dt^{q}$ follows.

Equation (\ref{equation62}) imitates Eq.(\ref{equ16}) for the column matrix $\hat{\rho}^{(0)}(t)$ of one variable and may be transformed to a non homogeneous linear 
third-order differential equation:
\begin{align}\nonumber\label{equation66}
\dfrac{d^{3}}{d\tau^{3}}\hat{\rho}^{(0)}(\tau)-\dfrac{1}{\tau}\dfrac{d^{2}}{d\tau^{2}}\hat{\rho}^{(0)}(\tau)- 4\Big[\dfrac{2\lambda_{M}}{\tau}\hat{\rho}^{(0)}(\tau)\\&&\hspace{-5cm}-(\tau^{2}+2\lambda_{M})\dfrac{d}{d\tau}\hat{\rho}^{(0)}(\tau)\Big]=
-6\Delta^{2}\hat{\mathcal{X}}(\tau),
\end{align}
where
\begin{align}\nonumber\label{equation67}
 &&\hspace{-1.5cm}\hat{\mathcal{X}}(\tau)=\dfrac{1}{\sqrt{\alpha}}\dfrac{d^{2}}{d\tau^{2}}\hat{\mathcal{J}}(\tau)\hat{\mathcal{Q}}^{(0)}(\tau)
- \dfrac{1}{\sqrt{\alpha}}\dfrac{d}{d\tau}\hat{\mathcal{J}}(\tau)\hat{\mathcal{Q}}^{(0)}(\tau)\\&&\hspace{-5.3cm}+\dfrac{4\tau^{2}}{\sqrt{\alpha}}\hat{\mathcal{J}}(\tau)\hat{\mathcal{Q}}^{(0)}(\tau),
\end{align}with $\lambda_{M}=\Delta^{2}_{M}/2\alpha$ being the LZ parameter in matrix form.
The physical sense of $\hat{\mathcal{X}}(\tau)$ may be achieved by expressing
$\hat{\rho}_{12}^{(0)}(\tau)=C_{1}^{(1)}(\tau) C_{2}^{(1)*}(\tau)$ and $\hat{\rho}_{21}^{(0)}(\tau)=C_{1}^{(1)*}(\tau)C_{2}^{(1)}(\tau)$
through Weber's functions.

Considering Eq.(\ref{equation64}), the solution of Eq.(\ref{equation66}) may be obtained with the aid of  $\hat{\rho}^{(0)}(-\infty)=-[1\hspace{.2cm}1]^{\textmd{T}}$ and the following conditions
\begin{align}\label{equation68}
\dfrac{d^{2}}{d\tau^{2}}\hat{\rho}^{(0)}(\tau)\mid_{\tau=-\infty}=-8\lambda_{M}, \quad \dfrac{d}{d\tau}\hat{\rho}^{(0)}(\tau)\mid_{\tau=-\infty}=0.
\end{align}
Then the solution of (\ref{equation66}) is given by
\begin{align}\nonumber\label{equation69}
&&\hspace{-.5cm}\hat{\rho}_{+}^{(0)}(\tau)= \lambda e^{-\pi \lambda}\Big( 2\lvert D_{-i\lambda }(-i\mu_{0} \tau)\rvert^{2} \lvert D_{-i\lambda-1}(-i\mu_{0} \tau)\rvert^{2}
 \\&&\hspace{-7cm}+\lambda\Big[\lvert D_{-i\lambda-1}(-i\mu_{0} \tau)\rvert^{2}- \dfrac{1}{\lambda}\lvert D_{-i\lambda}(-i\mu_{0} \tau)\rvert^{2} \Big]^{2}\Big).
\end{align}
The nontrivial dynamics of the population difference for the three-level LZ transition at any given time $\tau$ is governed by Eq.(\ref{equation66}). Considering the limit $\tau \to \infty$, we have the population difference
\begin{align}\label{equation70}
\hat{\rho}_{+}^{(0)}(\infty)=\hat{\rho}_{-}^{(0)}(\infty)=12 e^{-3\pi \lambda}\sinh(\pi \lambda)-1.
\end{align}
Using Eq.(\ref{equation70}) and the conservation of probability $\textmd{Tr}\hat{\tilde{\rho}}(\tau)=1$, we arrive at the transition probabilities in the second part of TABLE.\ref{TAB1}.

\subsection{ Fast noise, spin-$1$}\label{Sec5.1}
For the proper apprehension of the reader we review briefly the effects of fast noise on a three-level system. In the spirit of previous derivations, we transform Eqs.(\ref{equ58})-(\ref{equ59}) to
\begin{align}\label{equation71}
\dfrac{d}{dt}\hat{\rho}(t)=-4\int_{-\infty}^{t}\cos\Big[\alpha(t^{2}-t_{1}^{2})\Big]\hat{\mathcal{R}}_{M}(\left|t-t_{1}\right|)
\hat{\rho}(t_{1})
dt_{1}.
\end{align}
Here,
\begin{align}\label{equation72}
\hat{\rho}(t)=
\begin{bmatrix}
\langle \hat{\rho}_{+}(t)\rangle\\
\langle \hat{\rho}_{-} (t) \rangle
\end{bmatrix},
\end{align}
and
\begin{align}\label{equation73}
\hat{\mathcal{R}}_{M}(\left|t-t_{1}\right|)=2\begin{bmatrix}
1 & 1/2\\ 1/2 & 1
\end{bmatrix}
\hat{\mathcal{R}}(\lvert t-t_{1}\rvert)
\end{align}
the matrix correlator. Equation (\ref{equation71}) is structurally identical to Eq.(\ref{equ24}). Similarly we define  $\hat{\Omega}_{M}(t)=\hat{\Omega}^{(+)}_{M}(t)+\hat{\Omega}^{(-)}_{M}(t)$ as
\begin{align}\label{equation74}
\hat{\Omega}^{(\pm)}_{M}(t)=\int_{-\infty}^{\infty}\exp\Big[\pm i\tilde{\omega}(t) \xi\Big]\hat{\mathcal{R}}_{M}(\lvert \xi \rvert)d\xi.
\end{align}
The solution of Eq.(5.15) can be found as
\begin{widetext}
\begin{align}\label{equ65}
\langle \hat{\rho}_{11}(\infty)\rangle=\dfrac{1}{3}\Big(1+
\langle \hat{\rho}_{+}(-\infty)\rangle\Big[\sinh\dfrac{\theta}{2}+2\cosh\dfrac{\theta}{2}\Big]e^{-\theta}-\langle \hat{\rho}_{-}(-\infty)\rangle
\Big[\cosh\dfrac{\theta}{2}+2\sinh\dfrac{\theta}{2}\Big]e^{-\theta}\Big),
\end{align}
\begin{align}\label{equ66}
\langle \hat{\rho}_{00}(\infty)\rangle=\dfrac{1}{3}\Big(1-\langle \hat{\rho}_{+}(-\infty)\rangle
\Big[\cosh\dfrac{\theta}{2}-\sinh\dfrac{\theta}{2}\Big]e^{-\theta}+\langle \hat{\rho}_{-}(-\infty)\rangle
\Big[\sinh\dfrac{\theta}{2}-\cosh\dfrac{\theta}{2}\Big]e^{-\theta}\Big),
\end{align}
\begin{align}\label{equ67}
\langle \hat{\rho}_{22}(\infty)\rangle=\dfrac{1}{3}\Big(1-\langle \hat{\rho}_{+}(-\infty)\rangle
\Big[\cosh\dfrac{\theta}{2}+2\sinh\dfrac{\theta}{2}\Big]e^{-\theta}+\langle \hat{\rho}_{-}(-\infty)\rangle
\Big[\sinh\dfrac{\theta}{2}+2\cosh\dfrac{\theta}{2}\Big]e^{-\theta}\Big).
\end{align}
\end{widetext}
We considered the matrix transformation
\begin{align}\label{equation75}
\exp\Big[-\int_{-\infty}^{\infty}\hat{\Omega}_{M}(t')dt'\Big]
=e^{-\theta}
\begin{pmatrix}
\cosh\dfrac{\theta}{2} & -\sinh\dfrac{\theta}{2}\\ -\sinh\dfrac{\theta}{2} &  \cosh\dfrac{\theta}{2}
\end{pmatrix}.
\end{align}
The results agree with those of Pokrovsky\cite{pok}. The general form of these equations for arbitrary $t$ can be obtained by
$\theta \to \theta(t)$ (see  TABLE.\ref{TAB2}), where $\theta(t)$ is defined similarly as in (\ref{equa31a}). On TABLE.\ref{TAB2}, we show
infinite time transition probabilities for all possible initial occupation of the system.
\begin{table}{}
\begin{tabular}{lcr}
\hline
\hline
Initial occupation for $t=-\infty$ & $\rvert$ Final occupation for $t=\infty$ \\
\hline\\
1 & $ \frac{1}{3}(1+\frac{3}{2}e^{-\theta/2}+\frac{1}{2}e^{-3\theta/2}) $ \\ \\ 0 & $\frac{1}{3}(1-e^{-3\theta/2})$\\ \\ 0 & $\frac{1}{3}(1-\frac{3}{2}e^{-\theta/2}+\frac{1}{2}e^{-3\theta/2} )$\\
\\ \\
0 & $ \frac{1}{3}(1-e^{-3\theta/2}) $ \\ \\ 1 & $\frac{1}{3}(1+2e^{-3\theta/2})$\\ \\ 0 & $\frac{1}{3}(1-e^{-3\theta/2})$\\
\\ \\
0 & $\frac{1}{3}(1-\frac{3}{2}e^{-\theta/2}+\frac{1}{2}e^{-3\theta/2} ) $ \\ \\ 0 & $\frac{1}{3}(1-e^{-3\theta/2})$\\ \\ 1 & $\frac{1}{3}(1+\frac{3}{2}e^{-\theta/2}+\frac{1}{2}e^{-3\theta/2} )$\\
\hline
\hline
\end{tabular}
\caption{{\small Fast noise transition probabilities in the three-level system.}}\label{TAB2}
\end{table}
One can see that the transition probabilities for $S=1$ have the same form as for $S=1/2$. In the white noise approximation, we have the same probability distribution for all the triplet states.
\subsubsection*{Spin-$1$ in a constant off-diagonal field and a fast transverse random field}\label{Sec5.1.1}
We investigate the LZ transition assisted by fast-noise. The two-component noise is defined by Eq.(\ref{equa41}).
The mean-value of the stochastic function describing noise in $X$-direction is non-zero:
\begin{align}\label{equ68a}
\langle\Theta_{\pm}(t)\Theta_{\pm}(t_{1})\rangle\neq0.
\end{align}
The matrix density describing the noise assisted transition may now be represented as follows:
\begin{align}\label{equ68}
\hat{\rho}^{(\textmd{SF})}=
\begin{bmatrix}
 \langle\hat{\rho}_{+}^{(\textmd{SF})}(t)\rangle\\
 \langle\hat{\rho}_{-}^{(\textmd{SF})}(t)\rangle
\end{bmatrix}
\quad \textmd{and} \quad
\mathcal{Q}^{(\textmd{SF})}(t)=
\begin{bmatrix}
 \langle\hat{\rho}_{21}^{(\textmd{SF})}(t)\rangle\\
 \langle\hat{\rho}_{12}^{(\textmd{SF})}(t)\rangle
\end{bmatrix}.
\end{align}
The dynamics of the system is described by the equation:
\begin{align}\label{equ69}
\nonumber\dfrac{d}{dt}\hat{\rho}^{(\textmd{SF})}(t)
=-4\Delta^{2}_{M}\int_{-\infty}^{t}\cos\Big[\alpha(t^{2}-t_{1}^{2})\Big]\hat{\rho}^{(\textmd{SF})}(t_{1})dt_{1}
\\&&\hspace{-9cm}\nonumber-4\int_{-\infty}^{t}\cos\Big[\alpha(t^{2}-t_{1}^{2})\Big]\hat{\mathcal{R}}_{M}(\lvert t-t_{1}\rvert)
\hat{\rho}^{(\textmd{SF})}(t_{1})\\-6\Delta^{2}\hat{\mathcal{J}}(t)\hat{\mathcal{Q}}^{(\textmd{SF})}(t)
dt_{1}.
\end{align}
Considering equations (\ref{equa43}), (\ref{equa45}) then this permits us to write the solution of Eq.(\ref{equ69}).
We consider the decaying factors inducing dephasing that enter the final transition probabilities:
\begin{align}\label{equ70}
\hat{\rho}^{(\textmd{SF})}(t)=\exp\Big[-\int_{-\infty}^{t}\hat{\Omega}_{M}(t')dt'\Big]\hat{\rho}^{(0)}(t),
\end{align}
and
\begin{align}\label{equ71}
\hat{\mathcal{Q}}^{(\textmd{SF})}(t)=\exp\Big[-\int_{-\infty}^{t}\hat{\Omega}_{M}(t')dt'\Big]\hat{\mathcal{Q}}^{(0)}(t),
\end{align}
Equations (\ref{equ70}) and (\ref{equ71}) permit us to have the following relations:
\begin{widetext}
\begin{align}\label{equ72}
\langle \hat{\rho}_{11}^{(\textmd{SF})}(\infty)\rangle=\dfrac{1}{3}\Big(1+
\hat{\rho}_{+}^{(0)}(\infty)\Big[\sinh\dfrac{\theta}{2}+2\cosh\dfrac{\theta}{2}\Big]e^{-\theta}-\hat{\rho}_{-}^{(0)}(\infty)
\Big[\cosh\dfrac{\theta}{2}+2\sinh\dfrac{\theta}{2}\Big]e^{-\theta}\Big),
\end{align}
\begin{align}\label{equ72a}
\langle \hat{\rho}_{00}^{(\textmd{SF})}(\infty)\rangle=\dfrac{1}{3}\Big(1-\hat{\rho}_{+}^{(0)}(\infty)
\Big[\cosh\dfrac{\theta}{2}-\sinh\dfrac{\theta}{2}\Big]e^{-\theta}+\hat{\rho}_{-}^{(0)}(\infty)
\Big[\sinh\dfrac{\theta}{2}-\cosh\dfrac{\theta}{2}\Big]e^{-\theta}\Big),
\end{align}
\begin{align}\label{equ72b}
\langle \hat{\rho}_{22}^{(\textmd{SF})}(\infty)\rangle=\dfrac{1}{3}\Big(1-\hat{\rho}_{+}^{(0)}(\infty)
\Big[\cosh\dfrac{\theta}{2}+2\sinh\dfrac{\theta}{2}\Big]e^{-\theta}+\hat{\rho}_{-}^{(0)}(\infty)
\Big[\sinh\dfrac{\theta}{2}+2\cosh\dfrac{\theta}{2}\Big]e^{-\theta}\Big).
\end{align}

\begin{table*}{}
\begin{tabular}{lcr}
\hline
\hline
Initial occupation for $t=-\infty$ & $\rvert$ Final occupation for $t=\infty$ \\
\hline\\
1 & $\frac{1}{3}(1-[\frac{3}{2}e^{-\theta/2}-\frac{1}{2}e^{-3\theta/2}][1-4 e^{-2\pi\lambda}+3 e^{-4\pi\lambda}]+[\frac{3}{2}e^{-\theta/2}+\frac{1}{2}e^{-3\theta/2}][3e^{-4\pi\lambda}-2e^{-2\pi\lambda}])$
\\ \\ 0 & $\frac{1}{3}(1-e^{-3\theta/2}[1-4 e^{-2\pi\lambda}+3 e^{-4\pi\lambda}]-e^{-3\theta/2}[3e^{-4\pi\lambda}-2e^{-2\pi\lambda}])$
\\ \\ 0 & $\frac{1}{3}(1+[\frac{3}{2}e^{-\theta/2}+\frac{1}{2}e^{-3\theta/2}][1-4 e^{-2\pi\lambda}+3 e^{-4\pi\lambda}]-[\frac{3}{2}e^{-\theta/2}-\frac{1}{2}e^{-3\theta/2}][3e^{-4\pi\lambda}-2e^{-2\pi\lambda}])$\\
\\ \\
0 & $\frac{1}{3}(1+e^{-3\theta/2}[6e^{-2\pi\lambda}-6e^{-4\pi\lambda}-1])$ \\ \\ 1 &  $\frac{1}{3}(1-2e^{-3\theta/2}[6e^{-2\pi\lambda}-6e^{-4\pi\lambda}-1])$
\\ \\ 0 & $\frac{1}{3}(1+e^{-3\theta/2}[6e^{-2\pi\lambda}-6e^{-4\pi\lambda}-1])$\\
\\ \\
0 & $\frac{1}{3}(1+[\frac{3}{2}e^{-\theta/2}+\frac{1}{2}e^{-3\theta/2}][1-4 e^{-2\pi\lambda}+3 e^{-4\pi\lambda}]-[\frac{3}{2}e^{-\theta/2}-\frac{1}{2}e^{-3\theta/2}][3e^{-4\pi\lambda}-2e^{-2\pi\lambda}])$
\\ \\ 0 & $\frac{1}{3}(1-e^{-3\theta/2}[1-4 e^{-2\pi\lambda}+3 e^{-4\pi\lambda}]-e^{-3\theta/2}[3e^{-4\pi\lambda}-2e^{-2\pi\lambda}])$
\\ \\ 1 & $\frac{1}{3}(1-[\frac{3}{2}e^{-\theta/2}-\frac{1}{2}e^{-3\theta/2}][1-4 e^{-2\pi\lambda}+3 e^{-4\pi\lambda}]+[\frac{3}{2}e^{-\theta/2}+\frac{1}{2}e^{-3\theta/2}][3e^{-4\pi\lambda}-2e^{-2\pi\lambda}])$\\\\
\hline
\hline
\end{tabular}
\caption{{\small Transition probabilities for the three-level system in both a constant magnetic field and a classical transverse noise.}}\label{TAB3}
\end{table*}
\end{widetext}

From equations (\ref{equ72}) and (\ref{equ72b}) considering the dynamics of the system from an initial occupation for $t=-\infty$ to a final occupation, for $t=\infty$  this permit to write TABLE.\ref{TAB3} of the transition probabilities.

The quantities $\hat{\rho}_{+}^{(0)}(\infty)$ and $\hat{\rho}_{-}^{(0)}(\infty)$ can be obtained from TABLE.\ref{TAB1}. The corresponding transition probabilities are reported in TABLE.\ref{TAB3}.

\subsection{Slow noise, spin-$1$}\label{Sec5.2}

The transition probabilities for $S=1$  subjected to slow noise are obtained in the same spirit as
was discussed for $S=1/2$.
In a given realization $Q$ of noise,
the system of equations for the population differences (Eq.(\ref{equation60})-Eq.(\ref{equation61}))
 is reduced to Eq. (\ref{equation66}). The solutions of this problem for $\tau=\infty$ are  derived via Eq.(\ref{equation70}).

\begin{figure}[!h]
  \begin{center}
    \leavevmode
    \subfloat[]{%
      \label{fig10a}
      \includegraphics[width=4.3cm, height=35mm]{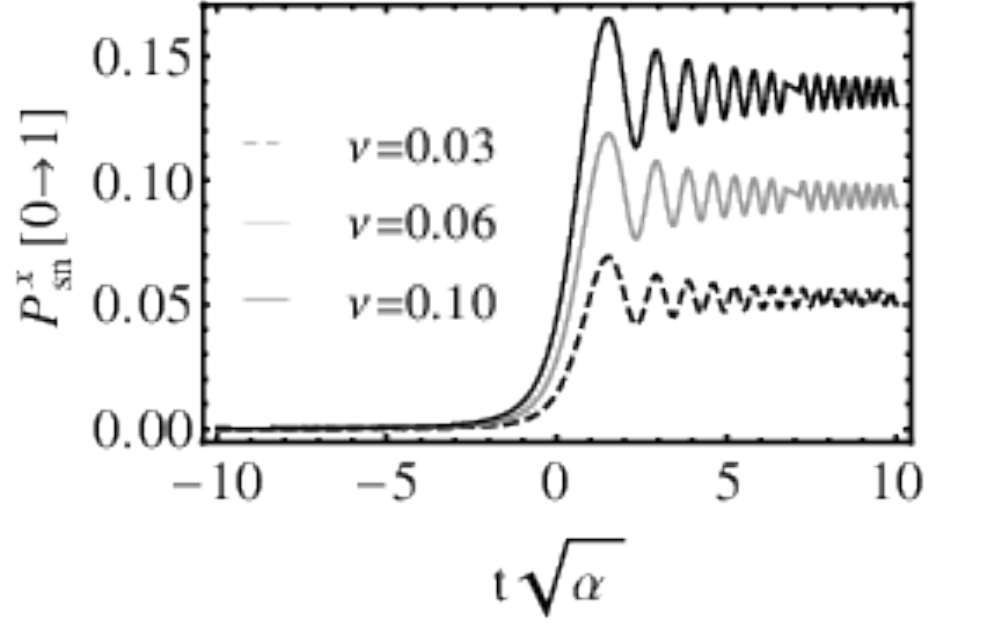}}
    \hspace{-4mm}
    \subfloat[]{%
      \label{fig10b}
      \includegraphics[width=4.3cm, height=35mm]{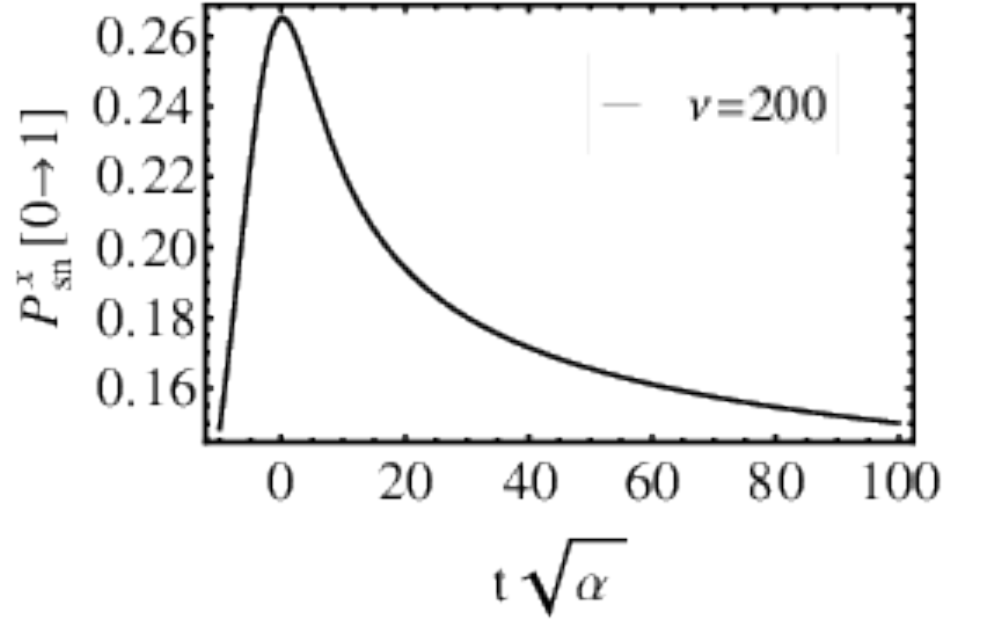}}
    \hspace{-4mm}
\subfloat[]{%
      \label{fig10c}
      \includegraphics[width=4.3cm, height=35mm]{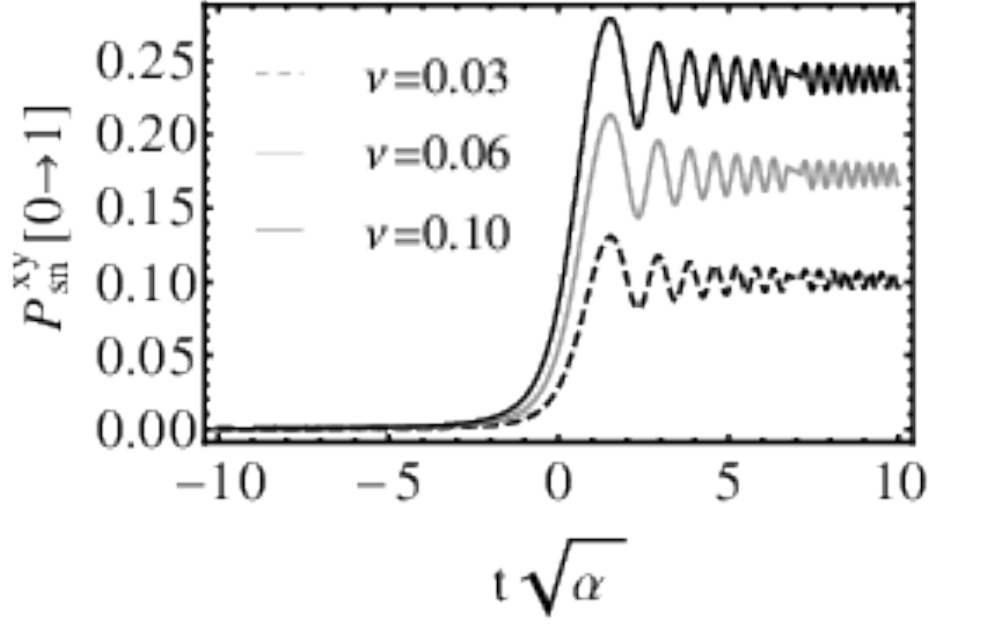}}
    \hspace{-4mm}
\subfloat[]{%
      \label{fig10d}
      \includegraphics[width=4.3cm, height=35mm]{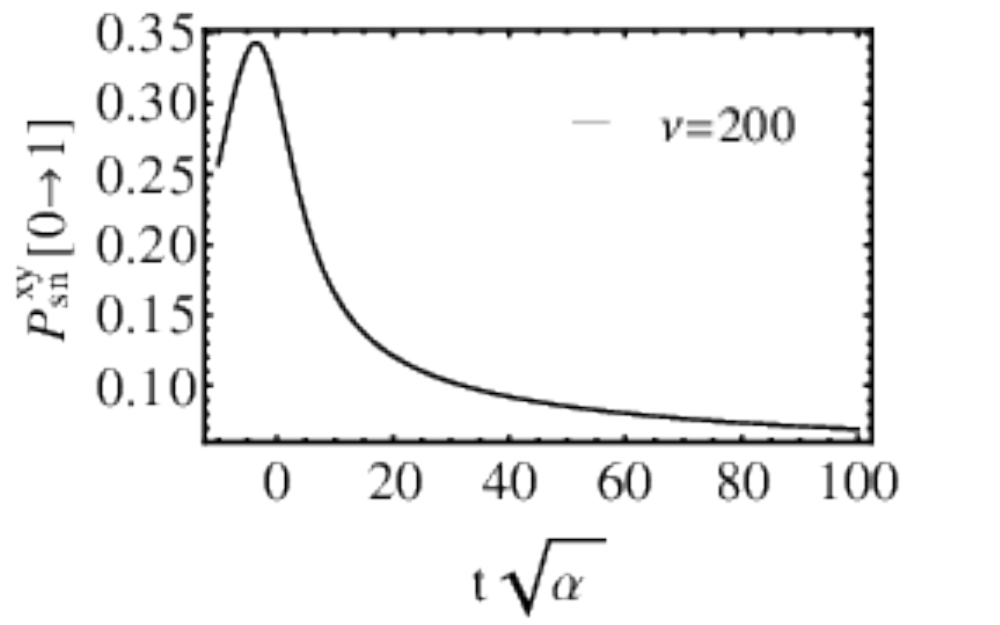}}
 \end{center}
\caption{Time evolution of the LZ transition probability in the diabatic basis of the three-level system in the presence of slow 
one- [ (a) and (b)] and two-component  [(c) and (d)] transverse noises (see discussion in the text). Panels (a) and (c) represent the results of numerical calculations for the small- amplitude noise. The data for the large-amplitude noise are shown on the panels (b) and (d).
} \label{FIG9}
\end{figure}

The LZ solutions for the case of one-component slow transverse noise are given by (see also Fig. \ref{FIG9})
\begin{align}\nonumber\label{equ73}
\mathcal{P}_{sn}^{x}[0\to1](t)=2\Big[\dfrac{\exp(-2\pi\lambda\Phi_{1}(t))}{\sqrt{1+\dfrac{2\pi\eta^{2}}{\alpha}\Big[F(t)+\ln W(t)\Big]}}
-\\&&\hspace{-4.5cm}\dfrac{\exp(-4\pi\lambda\Phi_{2}(t))}{\sqrt{1+\dfrac{4\pi\eta^{2}}{\alpha}\Big[F(t)+\ln W(t)\Big]}}\Big],
\end{align}
\begin{align}\nonumber\label{equ74}
\mathcal{P}_{sn}^{x}[0\to0](t)=1-\dfrac{4\exp(-2\pi\lambda\Phi_{1}(t))}{\sqrt{1+\dfrac{2\pi\eta^{2}}{\alpha}\Big[F(t)+\ln W(t)\Big]}}
+\\&&\hspace{-4.5cm}\dfrac{4\exp(-4\pi\lambda\Phi_{2}(t))}{\sqrt{1+\dfrac{4\pi\eta^{2}}{\alpha}\Big[F(t)+\ln W(t)\Big]}}.
\end{align}
For the two-component transverse noise, the transition probabilities read:
\begin{align}\nonumber\label{equ76}
\mathcal{P}_{sn}^{xy}[0\to1](t)=2\Big[\dfrac{\exp(-2\pi\lambda\Phi_{1}(t))}{1+\dfrac{2\pi\eta^{2}}{\alpha}\Big[F(t)+\ln W(t)\Big]}
-\\&&\hspace{-4cm}\dfrac{\exp(-4\pi\lambda\Phi_{2}(t))}{1+\dfrac{4\pi\eta^{2}}{\alpha}\Big[F(t)+\ln W(t)\Big]}\Big],
\end{align}
and
\begin{align}\nonumber\label{equ77}
\mathcal{P}_{sn}^{xy}[0\to0](t)=1-\dfrac{4\exp(-2\pi\lambda\Phi_{1}(t))}{1+\dfrac{2\pi\eta^{2}}{\alpha}\Big[F(t)+\ln W(t)\Big]}
+\\&&\hspace{-4cm}\dfrac{4\exp(-4\pi\lambda\Phi_{2}(t))}{1+\dfrac{4\pi\eta^{2}}{\alpha}\Big[F(t)+\ln W(t)\Big]}.
\end{align}

In Eqs.(\ref{equ73})-(\ref{equ77}) the notation $a \to b$ denotes the transition from the diabatic 
state $\rvert a \rangle$ to state $\rvert b \rangle$. $\mathcal{P}_{sn}^{\ell}[0\to1]=\mathcal{P}_{sn}^{\ell}[0\to2],$ with $\ell=x, xy$. 
Solutions (\ref{equ73})-(\ref{equ77}) represent the general LZ transition probabilities for a three-level system in the presence of the slow 
one- and two- dimensional transverse noise. This is also relevant for both noise-induced and noise-assisted transitions. Letting $\lambda=0$, 
in Eqs.(\ref{equ73})-(\ref{equ77}) we achieve a slow- noise-induced LZ transition.


\section{Discussion on effects of noise on Landau Zener times}\label{Sec6}

It is well-known that if one deals with a system of consequent Landau-Zener transitions,
it is not sufficient to characterize a behavior of such system by asymptotic values of probabilities.
One also needs to define a tunnel time\cite{Vitanov1999, Yan2012, Buttiker, Burkard, Dykman, Gefen, Ben} in order to put a borderline between two cases when the consequent tunnel processes can or can not be considered independently. 

There exist several ways to define the tunnel Landau-Zener time for two level system. Although we are not going to dwell onto a detailed discussion of tunnel times in this paper, let us list
a few physical definitions. One possible approach is the so-called "internal clock" definition.
It is based on analysis of LZ probability behavior at finite times. As it has been pointed out several times along our discussion, 
the finite time probability dynamics is characterized by monotonous function
for slow adiabatic passage, while for sudden (rapid) transition it oscillates before saturation
at constant value. These oscillations correspond to interference processes 
and determine the population of two states. Therefore, the "internal clock" approach defines
the Landau-Zener time as the width of transition to its asymptotic value (see
Refs.[\onlinecite{Gefen, Ben}] for detailed discussion).

An alternative approach to a definition of LZ times is based on "external clock" probe. In that case,
the LZ Hamiltonian is perturbed by a periodic transverse field $\delta\hat{\mathcal{H}}(t)=\epsilon \sin(\omega t+\phi)$, where, $\omega$ is the frequency 
of the field and $\phi$ is its initial phase.
The LZ time is determined through analysis of infinite time probability as a function of external field frequency (see details in Ref. \onlinecite{Ben})

Both definitions consistently lead to estimation of LZ times as $\tau_{\textmd{LZ}}=\Delta/\alpha$ for slow adiabatic passage and $\tau_{\textmd{LZ}}=1/\sqrt{\alpha}$ for rapid passage. Obviously, both definitions
can be straightforwardly generalized for multi-level LZ transitions.
 
Let us consider a slow noise as a special case of "external clock". We add a perturbation $\delta\hat{\mathcal{H}}=2f_{x}(t)S^{x}$ to the system such that the coupling $\Delta$ is deviated as
 $\tilde{\Delta}(t)=\Delta+f_{x}(t)$. This case has been discussed in the Sec. \ref{Sec3.2.1} and corresponds to a non-centered  one-component transverse noise. 
The role of noise is to frustrate the spins in the direction of the Zeeman field. 
Let us consider a square fluctuation of the Bloch's vector as a probe for LZ time:
\begin{align}\label{equ87}
\langle(\delta \vec{b})^{2}\rangle=\langle\vec{b}^{2}\rangle-\langle \vec{b}\rangle^{2}.
\end{align}
Since the classical noise only dephases the system and does not create any dissipation in it, the condition $\vec{b}^{2}=1$ holds.
In general, $\langle(\delta \vec{b})^{2}\rangle$ mixes the diagonal and off-diagonal components of the density matrix 
but for the two- and three- level systems subject to classical transverse noise we consider thus far, $\langle b_{x}\rangle=\langle b_{y}\rangle=0$ and $b_{z}(t)=\hat{\rho}_{11}(t)-\hat{\rho}_{22}(t)$.
Thus we write,
\begin{align}\label{equ88}
\langle(\delta \vec{b})^{2}\rangle=4 \mathcal{P}_{sn}(t)(1-\mathcal{P}_{sn}(t)).
\end{align}
The subscript $sn$ refers as usual to slow noise. It should however be noted that relation (\ref{equ88}) works both for two- and three- level systems under the assumption that the system 
initially prepared in one of the (upper or lower) diabatic states. 

\begin{figure}[]
  \begin{center}
    \leavevmode
    \subfloat[]{%
      \label{fig14a}
      \includegraphics[width=4.3cm, height=35mm]{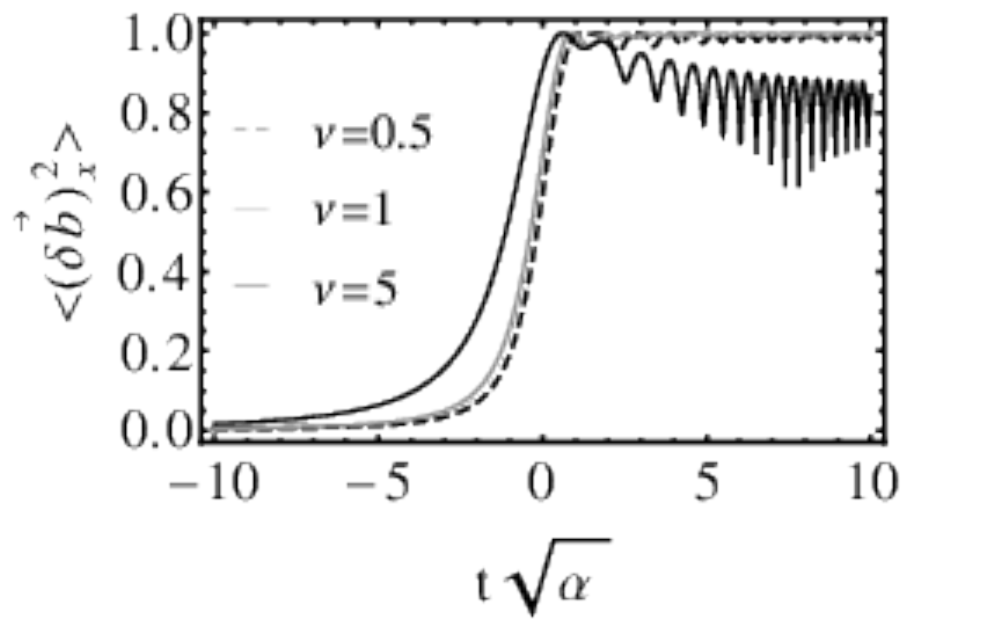}}
    \hspace{-6mm}
    \subfloat[]{%
      \label{fig14b}
      \includegraphics[width=4.3cm, height=35mm]{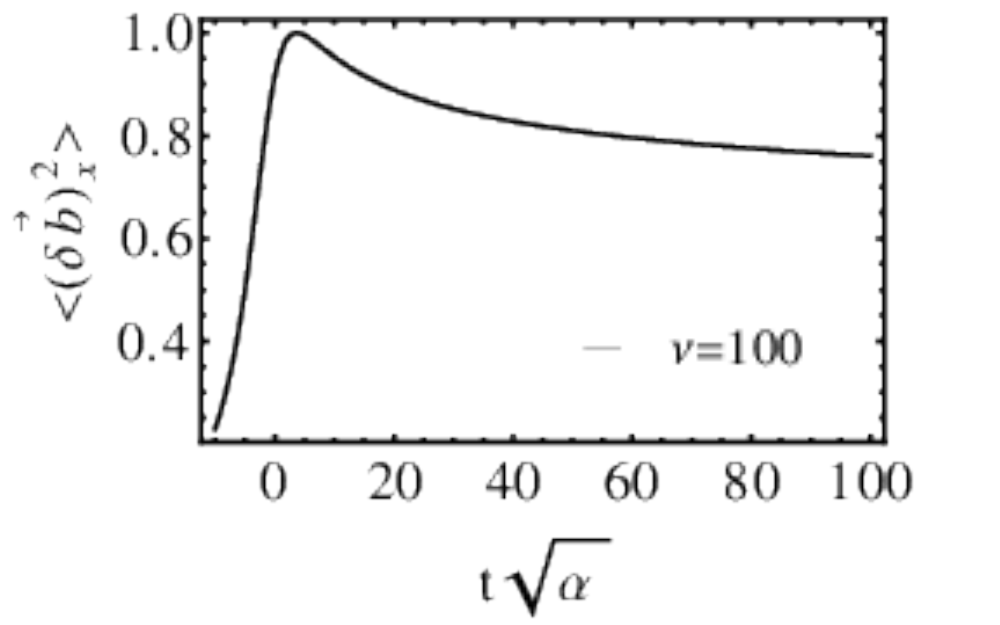}}
     \end{center}
 \caption{Typical time evolution of the fluctuation of the square of the Bloch vector given by 
 Eq.(\ref{equ88}). (a) The result for small and intermediate values of the noise amplitude. The large-amplitude noise results are presented in (b).} \label{FIG14}
\end{figure}

With these ideas in mind, we check the numerical behavior of $\langle(\delta \vec{b})^{2}\rangle$ for 
these initial conditions of the spin. Essential results are depicted by Fig.\ref{FIG14}. Interestingly, 
$\langle(\delta \vec{b})^{2}\rangle$ abruptly increases around the anticrossing region and saturates to its top value, 
confirming a spin flip transition. After the transition, the variance slightly fluctuates (slight decay of $\langle(\delta \vec{b})^{2}\rangle$) in the direction of the Zeeman field for adiabatic
 addition of noise (see Figs.\ref{fig14b}):
  \begin{align}\label{equ89}
\langle(\delta \vec{b})^{2}\rangle\leq\langle(\delta \vec{b})^{2}\rangle_{max}.
\end{align}
The two- and three- level systems,  seem no longer sensitive to any addition of noise from certain value of $\nu$ 
states are in thermal equilibrium. The transition time may then be defined as the 
particular moment when the square fluctuation of the Bloch's vector achieved its maximum value.

This qualitative definition can be experimentally probed. For the physical realization of this condition, we pose that
$\langle(\delta \vec{b})^{2}\rangle$ is a function of the variable $\mathcal{P}_{sn}(t)$.
 Then, the latter behaves as the quadratic function $h(x)=4x(1-x)$ defined in the real space. The mathematical requirement for a maximum of a function yields
\begin{align}\label{equ90}
\mathcal{P}_{sn}(\tau_{\textmd{LZ}})=\dfrac{1}{2}.
\end{align}
This definition actually coincides with the half-width condition
$\mathcal{P}_{sn}(\tau_{\textmd{LZ}})=\frac{1}{2}\mathcal{P}_{max}$,
where  $\mathcal{P}_{max}$  is the maximum value of probability. 
We emphasize in addition that the definition (\ref{equ90}) holds both for
two- and three- level systems.

\section{Conclusions}\label{Sec7}

In this paper, we discuss the effects of a transverse colored noise on both two and three-level systems subject to LZ transitions. The approximate solution for the traditional LZ problem is written down in terms of Fresnel's integrals  and appears to be useful for exploring the effects of slow noise at finite times. We demonstrated in the framework of von-Neumann equation that the effects of noise on a two-level system were regulated by an integral-differential master equation of the form (\ref{equ15}). We showed that for a fast
Gaussian noise it is sufficient to average that equation while for the slow noise, a correct procedure is based on averaging the solutions over the Gaussian realization of the noise. These arguments have been found to be general for the description of multilevel systems  where complicated interference patterns are expected.
The solution of Bloch's equation is generalized  for the finite-time LZ probabilities of two-and three- level system models in the presence of slow noise. We have essentially shown that for any
initial preparation of noise along one of transverse directions ($X$-noise) or two-component transverse  noise ($XY$-noise) the probability is renormalized by
new functions with shapes of standard LZ curves. The famous frequent exponentials appearing in LZ transition probabilities are considerably discriminated by an inverse square-root function of $\eta^{2}/\alpha$.

In the absence of noise, we showed that population difference for two- and three- level systems can be found as a solution of a third-order linear differential equation. 
The solution of this equation is given in terms of products of the parabolic cylinder Weber's functions.  We investigated solutions by evoking an isomorphism between Schr\"odinger's and Bloch's pictures.
In the presence of noise in general, the equations for density matrix elements are integral-differential
equations. Their solutions can be found through the averaging procedure discussed in the paper.

In conclusion, we would like to mention various realization of two- and three- level Landau-Zener transitions in recent quantum transport experiments \cite{Foletti2009, Foletti2010, Foletti2011}. The two-electron spin quantum bits are manipulated by the gate voltage applied to GaAs double quantum dot in the presence of external transverse magnetic field. 
The low-energy two-electron states in a double quantum well are given by three singlet and one triplet states. While the singlet states are not affected by the external magnetic field, the degeneracy of the triplet state is lifted out by the external Zeeman field.
In addition, there is a fluctuating Overhauser's field appearing due to a hyperfine interaction of electrons and  nuclear magnetic field of Ga and As sublattices of a host material. 
On one hand, the slowly fluctuating Overhauser's field is known to be responsible for both decoherence and dephasing \cite{KLG}.
On the other hand, if the double dot is not symmetric, the hyperfine magnetic field can result in transitions between singlet and triplet states. Therefore, in addition to three singlet states which form a three-level system, one of the triplet component should also be taken into account. Moreover, the transition between the singlet and triplet states provides a mechanism of nuclear spin polarization and 
effective cooling  the nuclear subsystem. Thus, the Overhauser's field leads to two competing effects of both nuclear polarization
and depolarization due to relaxation and dephasing. The model we discuss in the paper does not account
for the effects of relaxation only addressing the question of dephasing by classical fast- and slow Gaussian noises. Nevertheless, the competition between the polarizing (due to two- and three- state transitions) and depolarizing (due to the dephasing) effects is fully taken into account. The suppression of the LZ transition by the Overhauser's field fluctuations characterizes the effective temperature associated with noise and can
give a qualitative explanation for the nuclear spin depolarization mechanism.

\appendix
\section{PERTURBATIVE SOLUTION OF THE LANDAU-ZENER PROBLEM}\label{App1}

The integral-differential equation for conventional LZ problem (\ref{equ16}) in the absence of noise can be solved iteratively by setting the perturbative series expansion of $\hat{\rho}^{(0)}(\tau)$
via the parameter $\Delta^{2}/\alpha$:
\begin{align}\label{equ97}
 \hat{\rho}^{(0)}(\tau)=\sum_{k=0}^{\infty}\Big(-\dfrac{4\Delta^{2}}{\alpha}\Big)^{k}\hat{\rho}_{k}^{(0)}(\tau).
\end{align}
Where $\hat{\rho}_{0}^{(0)}(\tau)=1$ and ,
\begin{align}\label{equ98}
 \nonumber\hat{\rho}_{k}^{(0)}(\tau)=\int_{-\infty}^{\tau}d\tau_{1}\int_{-\infty}^{\tau_{1}}d\tau_{2}\cos[ \tau_{1}^{2}-\tau_{2}^{2})]\times...\\
...\times\int_{-\infty}^{\tau_{2k-2}}d\tau_{2k-1}\int_{-\infty}^{\tau_{2k-1}}d\tau_{2k}\cos[ \tau_{2k-1}^{2}-\tau_{2k}^{2}].
\end{align}
(See Ref. \onlinecite{integral} for details of calculation of sophisticated multiple integrals (\ref{equ98}) appearing in a classical-mechanical problem of a ball rolling on a Cornu spiral.)

In the presence of noise, we do $\Delta \to \eta$ in  Eq.(\ref{equ97}) and the function $\hat{\rho}_{k}^{(0)}(\tau)$ is modified  $\hat{\rho}_{k}(\tau)$:

\begin{align}\label{equ99}
\nonumber\hat{\rho}_{k}(\tau)=\int_{-\infty}^{\tau}d\tau_{1}\int_{-\infty}^{\tau_{1}}d\tau_{2}\cos[\tau_{1}^{2}-\tau_{2}^{2}]\times...\\\nonumber
...\times\int_{-\infty}^{\tau_{2k-2}}d\tau_{2k-1}\int_{-\infty}^{\tau_{2k-1}}d\tau_{2k}\cos[\tau_{2k-1}^{2}-\tau_{2k}^{2}]\times\nonumber
\\ \times F^{(k)}(\tau_{1},\tau_{2},...,\tau_{2k}),
\end{align}
where
\begin{align}\nonumber\label{equ100}
F^{(k)}(\tau_{1},\tau_{2},...,\tau_{2k})=
\\&&\hspace{-2.5cm}\eta^{-2k}\langle f_{+}(\tau_{1})f_{-}(\tau_{2})...f_{-}(\tau_{2k-1})f_{+}(\tau_{2k})\rangle.
\end{align}
To calculate the higher order correlation function, the Wick theorem is used. For the zero-mean random variables $f_{+}(\tau)$  and $f_{-}(\tau)$ this theorem suggests that:

\begin{align}\nonumber\label{equation100}
F^{(k)}(\tau_{1},\tau_{2},...,\tau_{2k})=\eta^{-2k}\\&&\hspace{-4cm}
 \left\{
  \begin{array}{lll}
   \sum_{pairs}\prod_{n=1}^{k}\langle f_{+}(\tau_{2n-1})f_{-}(\tau_{2n})\rangle, \quad \textmd{for even} \quad k,\\
    0, \quad \textmd{for odd} \quad k.
    \end{array}
\right..
\end{align}
The summation $\sum_{pairs}$  runs over all possible combinations of pairs out of the 2k variables $(\tau_{1},\tau_{2},...,\tau_{2k})$. Calculations for a one-component transverse noise lead to the Kayanuma result\cite{kay1985}
($\gamma_0=\gamma/\sqrt{\alpha}$):
\begin{align}\label{equa101}
F^{(k)}(\tau_{1},\tau_{2},...,\tau_{2k})=\sum_{pairs}\exp\Big(-\gamma_{0}\sum_{n=1}^{k}\lvert \tau_{2n-1}-\tau_{2n}\rvert\Big),
\end{align}
while for the two-component transverse noise ($XY$-noise)
\begin{align}\label{equa102}
F^{(k)}(\tau_{1},\tau_{2},...,\tau_{2k})=\sum_{pairs}2^{k}\exp\Big(-\gamma_{0}\sum_{n=1}^{k}\lvert \tau_{2n-1}-\tau_{2n}\rvert\Big).
\end{align}
For slow or fast noise, we do respectively $\gamma\rightarrow0$ and the former time-dependent function (\ref{equ100}) turns out to be a simple combinatorial factor. The long-time asymptotic value of $\hat{\rho}_{k}^{(0)}(\tau)$ becomes
\begin{align}\label{equa103}
\hat{\rho}_{k}^{(0)}(\infty)=\dfrac{\pi^{k}}{2^{2k-1}k!}, \quad k\ge1.
\end{align}
This helps to find (\ref{equ7}) in a perturbative expansion:
\begin{align}\label{equa104}
 P_{\textmd{LZ}}(\infty)=-\sum_{k=1}^{\infty}a_{k}\Big(\dfrac{\pi\Delta^{2}}{\alpha}\Big)^{k}.
\end{align}
Here, $a_{k}=(-1)^{k}/k!$ and the radius of convergence for Eq.(\ref{equa104}) equals infinity. Considering the limit of slow or fast noise,
the probability (\ref{equa104}) is modified by a coefficient in the perturbative expansion.  We write the solution of the integral equation (\ref{equ24})
for the cases of slow and fast noise driven LZ transition as
\begin{align}\label{equa105}
 P_{\textmd{LZ}}(\infty)=-\sum_{k=1}^{\infty}a_{k}b_{k}\Big(\dfrac{\pi\eta^{2}}{\alpha}\Big)^{k},
\end{align}
where $b_{k}$ are the combinatorial factors that appear after averaging over noise realization and depend on either fast or slow noise.

\subsection{Slow noise}\label{App1.1}

It has been shown in Ref. \onlinecite{Kiselev2009} that for slow noise, the combinatorial factor $b_{k}$ is
 expressed as:
\begin{align}\label{equa106}
 b_{k}=\sum_{pairs}1=(2k-1)!!,
\end{align}
for an $X$-noise model and
\begin{align}\label{equa107}
 b_{k}=\sum_{pairs}2^{k}=2^{k}k!,
\end{align}
for an $XY$-noise one.

\subsection{Fast noise}\label{App1.2}
The case of fast noise is completely different. In contrast with the slow noise, as shown in Ref. \onlinecite{kay1984} only a single term out of the $(2k-1)!!$ pairings in Eq.(\ref{equ100}) contributes to the summation for an X- noise:

\begin{align}\label{equa108}
b_{k}=2^{k-1}.
\end{align}
For an $XY$- model, noise contributes as
\begin{align}\label{equa109}
b_{k}=2^{2k-1}.
\end{align}

Equation (\ref{equa105}) can be viewed as a result of averaging the LZ probability over disorder noise realizations (the exponential function in LZ probability containing the fluctuating field is an ensemble average). This is in contrast to the fast noise case for which the argument of exponential function in the LZ probability is proportional to ''two-point noise correlation function'' (the argument is disorder average). Moreover, the statement concerning disorder averaging remains true for any finite time transition probability. Therefore, the same time dependent function $F(t)+\ln W(t)$ will enter the equation for the finite time slow noise driven LZ transition. Since the coefficient $b_{k}$ strongly depends on $k$, the radius of convergence of the perturbative expansion must also be found. After summing up the perturbative series for the LZ probability within the circle of convergence, the function has to be analytically continued into the outer part of the circle. 
We will identify analytical functions 
describing slow noise driven LZ probability and consider finite time LZ transition. Proceeding, with $b_{k}$ in Eqs.(\ref{equa106})-(\ref{equa109}) 
we will recover exactly all the LZ probabilities found previously.

\section{SPIN-$1$ DENSITY MATRIX EQUATIONS}\label{App2}
The equation of motion for the density matrix describing transitions in three-level systems can be obtained in the same way as we discussed in details earlier for $S=1/2$. 
In this appendix we present the full set of these equations for completeness of the discussion about connections between Schr\"odinger and Bloch pictures:

\begin{widetext}
\begin{align}\label{Equ1}
\dfrac{d\hat{\rho}_{+}(t)}{dt}=i\Theta_{+}(t)(2\hat{\rho}_{10}(t)-\hat{\rho}_{02}(t))-i\Theta_{-}(t)(2\hat{\rho}_{01}(t)-\hat{\rho}_{20}(t))
\end{align}
\begin{align}\label{Equ2}
\dfrac{d\hat{\rho}_{-}(t)}{dt}=-i\Theta_{+}(t)(2\hat{\rho}_{02}(t)-\hat{\rho}_{10}(t))+i\Theta_{-}(t)(2\hat{\rho}_{20}(t)-\hat{\rho}_{01}(t)).
\end{align}
Here,
 \begin{align}\label{Equ3}
\hat{\rho}_{10}(t)
=i\dfrac{\int_{t_{0}}^{t}\exp\Big(i\int_{t_{0}}^{t_{1}}\Theta^{z}(\tau')d\tau'\Big)\Theta_{-}(t_{1})\hat{\rho}_{+}(t_{1})dt_{1}}
{\exp\Big(i\int_{t_{0}}^{t}\Theta^{z}(\tau')d\tau'\Big)}+i\dfrac{\int_{t_{0}}^{t}\exp\Big(i\int_{t_{0}}^{t_{1}}\Theta^{z}(\tau')d\tau'\Big)\Theta_{+}(t_{1})\hat{\rho}_{12}(t_{1})dt_{1}}
{\exp\Big(i\int_{t_{0}}^{t}\Theta^{z}(\tau')d\tau'\Big)},
\end{align}
\begin{align}\label{Equ4}
\hat{\rho}_{12}(t)
=-i\dfrac{\int_{t_{0}}^{t}\exp\Big(2i\int_{t_{0}}^{t_{1}}\Theta^{z}(\tau')d\tau'\Big)\Theta_{-}(t_{1})(\hat{\rho}_{02}(t_{1})-\hat{\rho}_{10}(t_{1}))dt_{1}}
{\exp\Big(2i\int_{t_{0}}^{t}\Theta^{z}(\tau')d\tau'\Big)},
\end{align}
\begin{align}\label{Equ5}
\hat{\rho}_{02}(t)
=-i\dfrac{\int_{t_{0}}^{t}\exp\Big(i\int_{t_{0}}^{t_{1}}\Theta^{z}(\tau')d\tau'\Big)\Theta_{-}(t_{1})\hat{\rho}_{-}(t_{1})dt_{1}}
{\exp\Big(i\int_{t_{0}}^{t}\Theta^{z}(\tau')d\tau'\Big)}-i\dfrac{\int_{t_{0}}^{t}\exp\Big(i\int_{t_{0}}^{t_{1}}\Theta^{z}(\tau')d\tau'\Big)\Theta_{+}(t_{1})\hat{\rho}_{12}(t_{1})dt_{1}}
{\exp\Big(i\int_{t_{0}}^{t}\Theta^{z}(\tau')d\tau'\Big)}.
\end{align}
\end{widetext}
Here, $\hat{\rho}_{ij}(t)=\hat{\rho}^{*}_{ji}(t)$ and  $\hat{\rho}_{\pm}(t)=\hat{\rho}^{*}_{\pm}(t)$.


\section*{ACKNOWLEDGMENTS}
MBK work's has been supported by the Sandwich Training Educational Programme (STEP) of the Abdus Salam International 
Centre for Theoretical Physics (ICTP), Trieste, Italy. HNP work's has been supported through ICTP Postgraduate Diploma Programme. LCF appreciates visiting ICTP through Associates Scheme.  We are grateful to B. Altshuler, Ya. Blanter, G. Burkard, M. Dykman, Y. Gefen, S. Ludwig, F. Marquardt,
K. Matveev and A. Silva for fruitful discussions. We especially thank L. Levitov for drawing our 
attention to  Ref.[\onlinecite{Foletti2009}-\onlinecite{Foletti2011}] and V. Pokrovsky for detailed discussion of his works Ref.[\onlinecite{pok2003}-\onlinecite{pok}]. MNK appreciates discussions with K. Sengupta on various identities for 
the squares of parabolic cylinder functions \cite{Sengupta} and V. Gritsev for detailed introduction 
to the concept of dynamical symmetries.
We are thankful to Yu. Galperin and K. Kikoin for careful reading the manuscript and numerous useful and valuable suggestions.  MBK acknowledges comments of M. Tchoffo and K. Sadem on the manuscript.  MNK is grateful to KITP for hospitality. This research was supported in part by the National Science Foundation under Grant No. NSF PHY11-25915. 

\end{document}